\documentclass[
prb
,twocolumn
,amssymb, amsmath]{revtex4}
\usepackage[english]{babel}%
\usepackage{bm}%
\usepackage{graphicx}

\newcommand{\LCMO}{La$_{1-x}$Ca$_x$MnO$_3$}
\newcommand{\LSMO}{La$_{1-x}$Sr$_x$MnO$_3$}

\newcommand{\mean}[1]{\mathord{\left\langle #1 \right\rangle}}

\renewcommand{\>}{\mathord{\rangle}} 
\newcommand{\llangle}{\mathord{\langle\!\langle}}
\newcommand{\rrangle}{\mathord{\rangle\!\rangle}}

\newcommand{\cpa}{\mathord{c^+_{i\alpha\sigma}}}
\newcommand{\ca}{\mathord{c_{i\alpha\sigma}}}
\newcommand{\cpksa}{\mathord{c^+_{\mathbf k\alpha\sigma}}}
\newcommand{\cksa}{\mathord{c_{\mathbf k\alpha\sigma}}}

\newcommand{\halb}{\mathord{\frac{1}{2}}}
\newcommand{\dn}{\langle\Delta n\rangle}

\newcommand{\fk}{\mathord{\mathbf k}}

\newcommand{\fR}{\mathord{\mathbf R}}

\newcommand{\ek}{\mathord{\epsilon(\mathbf k)}}

\newcommand{\intmax}{\mathord{\int\limits^{+\infty}_{-\infty}}}

\begin{document}

\title{Magnetic Phase Diagrams of Manganites-like Local-Moment Systems with Jahn-Teller distortions}

\author{M. Stier}
\email{stier@physik.hu-berlin.de}
\author{W. Nolting}%
\affiliation{Festk\"orpertheorie, Institut f\"ur Physik, Humboldt-Universit\"at, 12489 Berlin, Germany}
\date{July 2008}%

\begin{abstract}
We use an extended two-band Kondo lattice model (KLM) to investigate the occurrence of different (anti-)ferromagnetic phases or phase separation depending on several model parameters. With regard to CMR-materials like the manganites we have added a Jahn-Teller term, direct antiferromagnetic coupling and Coulomb interaction to the KLM. The electronic properties are self-consistently calculated in an interpolating self-energy approach with no restriction to classical spins and going beyond mean-field treatments. Further on we do not have to limit the Hund's coupling to low or infinite values. Zero-temperature phase diagrams are presented for large parameter intervals. There are strong influences of the type of Coulomb interaction (intraband, interband) and of the important parameters (Hund's coupling, direct antiferromagnetic exchange, Jahn-Teller distortion), especially at intermediate couplings.
\end{abstract}

\maketitle

\section{Introduction}

The ferromagnetic Kondo lattice model, also known as double exchange or $sd$-model, is one of the basic models in solid state physics. It is valid for systems which can be divided into two subsystems. The first one describes the itinerant electrons and the other one localized electrons giving rise to finite permanent magnetic moments. A very prominent class of such materials are the manganites like \LSMO\ or \LCMO. Due to a crystal field splitting the five 3$d$ spin-up orbitals of the manganese ion are split into three $t_{2g}$ and two $e_g$ orbitals. The $t_{2g}$ spin-up states are fully occupied and provide the localized spin $S=\frac{3}{2}$ and in the $e_g$ states are $n=1-x$ itinerant electrons. But the KLM alone does not explain the complex phase diagrams of the manganites\cite{dag, martin,martin2} or other effects like the colossal magnetoresistance (CMR)\cite{milis}. Thus other physical effects seem to be important. This can be the superexchange which takes place between the $t_{2g}$ electrons and leads to a direct antiferromagnetic coupling $J_{AF}$. According to the Jahn-Teller theorem a crystal with degenerated states breaks its symmetry and therefore lowers the energy. This results in a splitting of the $e_g$ orbitals. Last but not least the electrons in the narrow $e_g$ bands experience a large Coulomb repulsion.\\
The large number of competing interactions leads to versatile impacts on the phase diagram\cite{dag_sc}. Not only that each effect influences the magnetic order for itself but they also act on each other. Thus it is necessary to understand these interactions and their mutual effects in detail. Theoretical work on this has already been done using different methods\cite{brink,peters,sheng,hotta,ryo,fishman}. \\
The composition of this paper is as follows: In the next section we will present our complete model. The methods we have used to solve this model are described in Sec. \ref{calc}. Afterwards we will show the numerical results in Sec. \ref{results}. There the magnetic phase diagrams will be presented which are derived by a comparison of the internal energies of the different phases. Finally we will come to a conclusion in Sec. \ref{concl}.

\section{\label{smod}Model}

We use a two-band ferromagnetic Kondo lattice model as the main part of our Hamiltonian.
\begin{equation}
 H_{\text{KLM}}=\sum_{\<i,j\>,\alpha,\sigma}T_{ij}\cpa c_{j\alpha\sigma}-J_H\sum_{i,\sigma,\alpha}\boldsymbol \sigma_{i\alpha}\cdot \mathbf S_{i}
\end{equation}
The first term describes the next-neighbor hopping of the itinerant electrons with the hopping matrix $T_{ij}$ and the fermion annihilation (creation) operators $c^{(+)}_{i\alpha\sigma}$ for electrons with spin $\sigma$ in the band $\alpha$ . An on-site coupling of the spin of the conduction electrons ($\boldsymbol\sigma_{i\alpha}$) to the spins of the localized electrons ($\mathbf S_i$) is done via the Hund's coupling $J_H$. The KLM for itself has a rich phase diagram\cite{john,chatt} and is often used as the only part to describe real materials\cite{sinj,cgd,wm} with some success. But special additions are needed to explain special features or more complex materials\cite{milis}. One of the most prominent interactions is the Coulomb repulsion represented by the Hubbard part
\begin{equation}
 H_U=U\sum_{i,\sigma \alpha}n_{i\alpha\sigma}n_{i\alpha\bar\sigma}+\sum_{i,\sigma,\sigma',\alpha}\tilde U^{\sigma\sigma'}n_{i\alpha\sigma}n_{i\bar\alpha\sigma'},
\end{equation}
where $n_{i\alpha\sigma}=\cpa\ca$ and a bar above an index means the opposite band or spin. If we choose $\tilde U^{\sigma\sigma'}=0$ there is only \textit{intra}band repulsion and with $\tilde U^{\sigma\sigma'}\neq0$ there is also \textit{inter}band repulsion. Besides this direct electron-electron interaction there can be a direct coupling of the localized spins.
\begin{equation}
 H_{AF} = J_{AF}\sum_{\langle i,j\rangle}\mathbf S_i\cdot\mathbf S_j \label{hss}
\end{equation}
This interaction is often generated by the superexchange of electrons. Therefore it is only used as an antiferromagnetic coupling ($J_{AF}>0$) in this paper. The last extension is due to the Jahn-Teller effect (JTE), which lowers the degeneracy of electron states by reducing the symmetry of the crystal. Mostly it is related to $3d$-orbitals, e.g. the $e_g$-orbitals in the manganites. In a standard notation it is written as
\begin{equation}
 H_{\text{JT}}=-g\sum_{i}(Q_{2i}T^x_i+Q_{3i}T^z_i)
+\halb k_{\text{JT}}(Q^2_{2i}+Q^2_{3i}).\label{jtham}
\end{equation}
The $Q_{2(3)i}$ are special JT modes and the $T^{z(x)}_i$ are pseudospin operators where the band index replaces the spin index of the usual spin operators. In our paper we set $k_{\text{JT}}=1$. With this final part we have the complete Hamiltonian
\begin{equation}
 \mathcal H = H_{\text{KLM}} +H_{U}+ H_{AF}+H_{JT}\label{model}
\end{equation}

\section{\label{calc}Calculation Methods}

To get magnetic phase diagrams at zero temperature we have to calculate the internal energy of the differently ordered magnetic configurations. We primarily focus on the ferromagnetic alignment and get the antiferromagnetic phases by a division of the whole chemical lattice into ferromagnetic sublattices.

\subsection{Ferromagnetic Phase}

For the internal energy we need the quasi-particle density of states $\rho_{\alpha\sigma}(E)$ (QDOS) which we can get from the one-particle Green's functions
\begin{eqnarray}
 \rho_{\alpha\sigma}(E)&=&-\frac{1}{\pi N}\sum_{\fk}\text{Im} G_{\fk\alpha\sigma}(E)\\
G_{\fk\alpha\sigma}(E)&=&\llangle \cksa;\cpksa\rrangle_E \nonumber\ .
\end{eqnarray}
In principle these Green's functions can be specified by solving the according equations of motion (EOM). Unfortunately there is no known exact analytical solution for this model thus we need approximation methods. Just looking at the first two parts in (\ref{model}) we can directly use the interpolating self-energy approach (ISA)\cite{isa} which has been successfully applied for the description of real materials (e.g. Ref.\cite{kreissl,niko,ms2}). This approach fulfills the exactly solvable limiting cases of the KLM (ferromagnetically ordered semiconductor, atomic limit, second order perturbation theory) and interpolates them by fitting free parameters via a high energy expansion. Therefore we expect reasonable results even between the limiting cases and it should hold for all band occupations, temperatures and all orders of Hund's coupling. Within this approach we get the self-energy
\begin{eqnarray}
\Sigma^{\text{ISA}}_{\alpha\sigma}(E)&=&-\halb J_H X_{\alpha,-\sigma}+ \label{self}\\
&+&\frac{1}{4}J_H^2\frac{a_{\alpha,-\sigma}G^{(0)}_{\alpha,-\sigma}(E-\halb z_{\sigma}J_H X_{\alpha,-\sigma})}{1-\halb J_H G^{(0)}_{\alpha,-\sigma}(E-\halb z_{\sigma}J_H X_{\alpha,-\sigma})}\nonumber\,
\end{eqnarray}
containing
\begin{eqnarray}
a_{\alpha\sigma}&=&S(S+1)-X_{\alpha\sigma}(X_{\alpha\sigma}+1)\nonumber\\
X_{\alpha\sigma}&=&\frac{\Delta_{\alpha\sigma}-z_{\sigma}\mean{S_z}}{1-\mean{n_{\alpha\sigma}}}\nonumber \\
\Delta_{\alpha\sigma}&=&\mean{S^{\sigma}_i c^+_{i\alpha,-\sigma}c_{\alpha\sigma}}+ z_{\sigma}\mean{S^z_i n_{i\alpha\sigma}}\nonumber\\
G^{(0)}_{\alpha\sigma}(E)&=&\frac{1}{N}\sum_{\fk}\frac{\hbar}{E+\mu-T_{\alpha\sigma}(\fk)}\nonumber\ .
\end{eqnarray}
In addition to the model parameters of (\ref{model}) we need $\Delta_{\alpha\sigma}$ and $\mean{n_{\alpha\sigma}}$ which can be self-consistently calculated via the spectral theorem from the full Green's function
\begin{equation}
G_{\fk\alpha\sigma}(E)=\hbar\frac{\gamma_{\alpha\sigma}}{E+\mu-T_{\alpha\sigma}(\fk)-\Sigma^{\text{ISA}}_{\sigma\alpha}(E)}\label{gf}
\end{equation}
via
\begin{eqnarray}
\mean{n_{\alpha\sigma}}& =&  -\frac{1}{\pi N} \sum_{\fk}\intmax dE\ f_-(E)\text{Im}G_{\fk\sigma\alpha}(E-\mu) \label{n_isa}\\
 \Delta_{\alpha\sigma}&=& -\frac{2}{\pi N J} \sum_{\fk}\intmax dE\ f_-(E)\times \nonumber\\
& &  \times[E-T_{\sigma\alpha}(\fk)]\text{Im}G_{\fk\sigma\alpha}(E-\mu)\label{d_isa}
\end{eqnarray}
with the Fermi function $f_-(E)$. The expectation value $\mean{S^z}$ has to be considered as an external parameter or can be taken from another method\cite{mrkky1}. In the upper Green's function the Hubbard part of the Hamiltonian is incorporated in the $\gamma_{\alpha\sigma}$ via an effectice medium approach.   Normally the Green's function (\ref{gf}) contains a second part (upper Hubbard band) which we left out choosing respectively $U\gg W,J$ or $\tilde U^{\sigma\sigma'}\gg W,J$. Thus (\ref{gf}) only describes electrons which have no Coulomb repulsion partner at their site. The probability that an electron has \textit{no} repulsion partner is
\begin{equation}
 \gamma_{\alpha\sigma} = \underbrace{1 - \mean{n_{\alpha,-\sigma}}}_{\text{intraband}}\underbrace{-\mean{n_{-\alpha,\sigma}}-\mean{n_{-\alpha,-\sigma}}}_{\text{interband}}\label{specw}\ .
\end{equation}
Since the $\gamma_{\alpha\sigma}$ are in the numerator of (\ref{gf}) they influence the spectral weight of the Green's function but they also act on the bandwidth in
\begin{equation}
 T_{\alpha\sigma}(\fk)=T^{(0)}_{\alpha}+\gamma_{\alpha\sigma}\left(\ek-T^{(0)}_{\alpha}\right) \label{tas}
\end{equation}
where $T^{(0)}_{\alpha}$ are the centers of gravity of the bands. For this reason we have a total spectral weight 
\begin{eqnarray}
\label{totsw} \Gamma = \sum_{\alpha\sigma} \gamma_{\alpha\sigma}=
	\begin{cases}
		4-n & \text{intraband}\\
		4-3n & \text{intra+interband}
	\end{cases}  
\end{eqnarray}
which depends on the electron density $n$. Especially for $n=1$ and interband repulsion we have completely filled bands ($\Gamma=n=1$) and therefore a Mott insulator.\\
We now have to add the Jahn-Teller Hamiltonian of Eq. (\ref{jtham}). This can be easily done when we treat the phonon operators classically and use a mean-field approximation\cite{dag,hotta2}. After performing the mean-field decoupling the phononic variables $Q_{2i},Q_{3i}$ are only coupled to mean values of the electrons and the ground state is defined by the mean values $\mean{Q_{2i}},\mean{Q_{3i}}$ due to the absence of quantum fluctuations. Introducing spherical coordinates $Q_2=Q\cos\theta$, $Q_3=\sin\theta$ and dressed operators 
\begin{eqnarray}
  c_{i\sigma\alpha=-1}&=& e^{i\theta/2}(\cos\frac{\theta}{2}c_{i\sigma,3z^2-r^2}+\sin\frac{\theta}{2}c_{i\sigma,x^2-y^2})\nonumber\\
c_{i\sigma\alpha=+1}&=& e^{i\theta/2}(\sin\frac{\theta}{2}c_{i\sigma,3z^2-r^2}-\cos\frac{\theta}{2}c_{i\sigma,x^2-y^2})\label{orb}
\end{eqnarray}
we can replace the two modes in (\ref{jtham}). These operators are superpositions of $3d$-orbitals in $z$-direction or in the $x$-$y$-plane, respectively. If we also assume translational invariance (i.e. a non-cooperative JTE) the JTE is not dependent on the angle $\theta$ but only on the magnitude of the distortion $Q$. In the ground state this distortion is defined by  $Q=g\mean{\Delta n}$ and we get the new form of the Hamiltonian (\ref{jtham})
\begin{equation}
 H_{JT}=\sum_{\fk\alpha\sigma}\left(z_{\alpha}g^2\dn\cpksa\cksa\ +\frac{1}{2}g^2\mean{\Delta n}^2\right)\label{jtham2},
\end{equation}
with $z_{\alpha=\pm 1}=\pm 1$ and the occupation difference
\begin{equation}
 \mean{\Delta n} = \sum_{\sigma} \left(\mean{n_{\alpha=-1,\sigma}} -\mean{n_{\alpha=+1,\sigma}}\right) \label{split}
\end{equation}
which has to be calculated self-consistently using (\ref{n_isa}). Since we want to add this term to the Green's function (\ref{gf}) we can neglect the second term in (\ref{jtham2}) because it plays no role in the equation of motion. But we have to keep it in mind if we calculate the internal energy. The first term means only a shift of the energy of the two band in different directions. Hence it can be incorporated in the centers of gravity $T^{(0)}_{\alpha}$ in (\ref{tas}) which have then to be calculated self-consistently, too.\\
Now we have found a solution for all parts of the Hamiltonian (\ref{model}) which contains electron operators and thus we can calculate the quasi-particle DOS $\rho_{\alpha\sigma}$. The treatment of the superexchange part (\ref{hss}) will be described in \ref{inten}. 

\subsection{Antiferromagnetic Phases}

\begin{figure}[tb] 
\includegraphics[width=\linewidth]{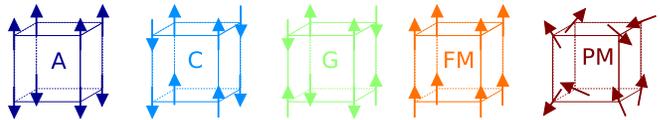}
\caption{\label{mag_types}(color online) The different types of magnetic order which are investigated in this paper. Arrows show the spin direction. A,C,G-type antiferromagnetism, FM/PM ferro-/paramagnetism.}
\end{figure}

In the section above we found a solution for the ferromagnetically ordered system. To extend this result to antiferromagnetic order we divide the whole chemical lattice into ferromagnetic sublattices. This means that we assume a N\'eel state which is known not to be the ground state due to quantum fluctuations. But at least for spins $S>\frac{1}{2}$ it should be close enough to the ground state to make reliable conclusions\cite{anderson,nolt_qdm}. For simplicity we only investigate types of antiferromagnetism which can be divided from a simple cubic lattice into two sublattices (Fig. \ref{mag_types}).
All of the electronic interaction parts in the Hamiltonian (\ref{model}) are local thus we have to sum over one more index $\nu$ which distinguishes the two sublattices. The hopping part needs two indices $\nu, \nu'$ since there is of course a hopping between different sublattices. Neglecting again $H_{AF}$ we get
\begin{equation}
 \mathcal H = H_s + \sum_{\nu}H^{\nu}_{\text{IA}}\ ,
\end{equation}
where $H^{\nu}_{\text{IA}}$ includes all electronic interaction parts (double exchange, Hubbard and JTE) in the magnetic sublattice $\nu$ and
\begin{equation}
 H_s=\sum_{\alpha,\sigma,\nu,\nu',\fk}\epsilon^{\nu\nu'}_{\fk}c_{\fk\alpha\sigma\nu}^{+}c_{\fk\alpha\sigma\nu'}\ .
\end{equation}
The dispersions $\epsilon^{\nu\nu'}_{\fk}$ are the Fourier-transformed hopping integrals
\begin{equation}
 \epsilon^{\nu\nu'}_{\fk} = \frac{1}{N}\sum_{\mean{i,j}}T_{ij}^{\nu\nu'}e^{-i\fk(\fR_i^{\nu}-\fR_j^{\nu'})}\ .
\end{equation}
We only allow hopping to next neighbors in the \textit{chemical} lattice. For solving the equation of motion (EOM) we define the Green's function
\begin{equation}
 G^{\nu\nu'}_{\fk\alpha\sigma}(E)=\llangle c_{\fk\alpha\sigma\nu};c_{\fk\alpha\sigma\nu'}^{+}\rrangle\ .
\end{equation}
To get the quasi-particle DOS $\rho^{\nu}_{\alpha\sigma}(E)$ we have to calculate the EOM for the $G^{\nu\nu}_{\fk\alpha\sigma}(E)$ which is
\begin{eqnarray}
 E G^{\nu\nu}_{\fk\alpha\sigma} &=&\hbar+\llangle[c_{\fk\alpha\sigma\nu},H_s];c_{\fk\alpha\sigma\nu}^+\rrangle+\\
&&+\llangle[c_{\fk\alpha\sigma\nu},H^{\nu}_{\text{IA}}];c_{\fk\alpha\sigma\nu}^+\rrangle
\nonumber\\ 
&=&\hbar + \epsilon^{\nu\nu}_{\fk}G^{\nu\nu}_{\fk\alpha\sigma}+\epsilon^{\nu\bar\nu}_{\fk}G^{\bar\nu\nu}_{\fk\alpha\sigma}+M^{\nu}_{\fk\alpha\sigma}G^{\nu\nu}_{\fk\alpha\sigma}\label{eom}
\end{eqnarray}
Here $[\dots]$ represents the commutator, $\bar\nu$ means the opposite sublattice. 
The higher Green's function is only affected by local interactions therefore we can use the sublattice self-energy $\llangle[c_{\fk\alpha\sigma\nu},H^{\nu}_{\text{IA}}];c_{\fk\alpha\sigma\nu}^+\rrangle = M^{\nu}_{\fk\alpha\sigma}(E) G^{\nu\nu}_{\fk\alpha\sigma}(E)$\footnote{Higher correlations due to sublattice mixing in the exact term are neglected. They appear in the next step of the EOM $ E \llangle[c_{\fk\alpha\sigma\nu},H^{\nu}_{\text{IA}}];c_{\fk\alpha\sigma\nu}^+\rrangle = \mean{[[c_{\fk\alpha\sigma\nu},H^{\nu}_{\text{IA}}],c_{\fk\alpha\sigma\nu}^+]_+ } + \llangle[[c_{\fk\alpha\sigma\nu},H^{\nu}_{\text{IA}}],\mathcal H];c_{\fk\alpha\sigma\nu}^+\rrangle$ with the full Hamiltonian $\mathcal H$.}.  Thus $M^{\nu}_{\fk\alpha\sigma}(E)$ is the self-energy of a pure ferromagnetic case comparable to (\ref{gf}). Now we need the Green's function $G^{\bar\nu\nu}_{\fk\alpha\sigma}(E)$ coming from the interlattice hopping. This leads to a second EOM
\begin{eqnarray}
E G^{\bar\nu\nu}_{\fk\alpha\sigma}
&=&\epsilon^{\bar\nu\nu}_{\fk}G^{\nu\nu}_{\fk\alpha\sigma}+\epsilon^{\bar\nu\bar\nu}_{\fk}G^{\bar\nu\nu}_{\fk\alpha\sigma}+M^{\bar\nu}_{\fk\alpha\sigma}G^{\bar\nu\nu}_{\fk\alpha\sigma} \label{eom3}
\end{eqnarray}
Other simplifications can be done if one considers some symmetries. The intra-sublattice hoppings should be the same for both sublattices because of the same chemical structure, i.e. $\epsilon^{\nu\nu}_{\fk}= \epsilon^{\bar\nu\bar\nu}_{\fk}=\tilde\epsilon_{\fk}$. This also holds for the inter-sublattice hoppings $\epsilon^{\bar\nu\nu}_{\fk}= \epsilon^{\nu\bar\nu}_{\fk}=t_{\fk}$. Since the two sublattices only differ by the spin direction we can replace the self-energy of the opposite sublattice by switching the spin 
\begin{equation}
M^{\nu}_{\fk\alpha\sigma}(E)=M^{\bar\nu}_{\fk\alpha\bar\sigma}(E)=M_{\fk\alpha\sigma}(E)\ .
\end{equation}
For the formal self-energy $M_{\fk\alpha\sigma}$ we now use the ISA self-energy plus the terms describing the Jahn-Teller effect and the Coulomb repulsion like in Eq. (\ref{gf}). With (\ref{eom}) and (\ref{eom3}) we find the full Green's function of one ferromagnetic sublattice
\begin{widetext}
 \begin{equation}
 G^{\nu\nu}_{\fk\alpha\sigma}(E)= G_{\fk\alpha\sigma}(E)=\hbar\frac{\gamma_{\alpha\sigma}}{E-\tilde T_{\alpha\sigma}(\fk)-\Sigma^{\text{ISA}}_{\alpha\sigma}(E)-|t_{\fk}|^2\left(E-\tilde T_{\alpha\bar\sigma}-\Sigma^{\text{ISA}}_{\alpha\bar\sigma}(E)\right)^{-1}}\label{gf2}\ ,
\end{equation}
\end{widetext}
where $\tilde T_{\alpha\sigma}(\fk) = T^{(0)}_{\alpha} +\gamma_{\alpha\sigma}(\tilde \epsilon_{\fk}-T^{(0)}_{\alpha})$ similar to (\ref{tas}). The Hubbard interaction is contained in the $\gamma_{\alpha\sigma}$ and the JT splitting in the $T^{(0)}_{\alpha}=z_{\alpha}\mean{\Delta n}g^2$ (cf. Eq. (\ref{jtham2})).
Although this treatment needed some approximations it fulfills the limiting cases $J_H\rightarrow 0$ and $\mean{S^z}\rightarrow 0$. In these cases the paramagnetic DOS is reproduced.

\subsection{Internal Energy\label{inten}}

To decide which magnetic phase is preferred at a special parameter set we have to compare the internal energies. In the KLM it is simply given as
\begin{equation}
 \frac{U'}{N}=\frac{\mean{H_{\text{KLM}}}}{N}=\sum_{\alpha\sigma} \intmax dE\ f_-(E) \rho_{\alpha\sigma}(E-\mu)E\ .
\end{equation}
It is a special feature of the KLM that some parts cancel each other out and the typical $\ek$-part is missing. The electronic parts of the Hamiltonian (\ref{model}) which do not belong to $H_{\text{KLM}}$ are contained in the Green's functions (\ref{gf}) and (\ref{gf2}) via an effective medium approach. Thus they are already taken into account by the quasi-particle DOS $\rho_{\alpha\sigma}(E)$. We just have to add the second term of (\ref{jtham2}) which had no influence on the EOMs but on the internal energy. This means for the electronic internal energy
\begin{equation}
 U_{\text{el}}/N = U'/N +\frac{1}{2}g^2\mean{\Delta n}^2.\label{uel}
\end{equation}
The only missing part is the energy of direct coupling between the localized moments (\ref{hss}). Since we are in the zero-temperature regime we use a mean-field decoupling
\begin{equation}
 \frac{U_{\text{AF}}}{N}=\frac{\mean{H_{\text{AF}}}}{N}= J_{AF}\mean{S^z}^2\left(a_{\text{p}}-a_{\text{ap}}\right) .\label{uss}
\end{equation}
We defined $a_{\text{(a)p}}$ as the number of (anti-)parallel aligned next-neighbor spins. This energy is lowest for highest antiferromagnetic order (G-type, all neighbors are antiparallel aligned) and vice versa highest for ferromagnetic order. The whole internal energy is now
\begin{equation}
 U/N=U_{\text{el}}/N+U_{\text{AF}}/N
\end{equation}

Ferromagnetic saturation is assumed for the respective (anti-)ferromagnetic phases. Thus we set $\mean{S^z}$ to the maximum value which is in the ISA\cite{isa} $\mean{S^z}_{\text{max}} = S\frac{S+1-n}{S+1}$. For the paramagnetic phase there is a vanishing mean value of the magnetization $\mean{S^z}=0$ and therefore also no contribution to the spin's internal energy (\ref{uss}) 

\subsection{Phase Separation \label{ps}}

\begin{figure}[b]
 	 \includegraphics[width=\linewidth]{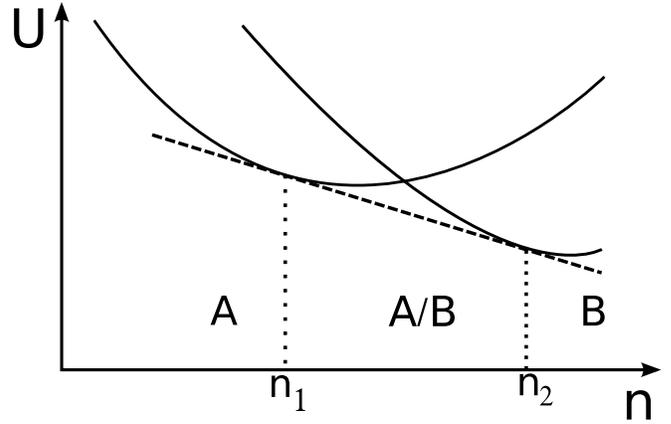}
	\caption{\label{ps_gen}Schematic view on the Maxwell construction. Between $n_1$ and $n_2$ the internal energy curves of the phases A and B are replaced by a straight line representing the occurrence of phase separation.}
\end{figure}
In the regions of phase separation the complete internal energy can be expressed by the internal energies of the phases $A,B$ as
\begin{equation}
 U^{PS} = (1-y) U^A(n_A) + y U^B(n_B)
\end{equation}
with the volume fraction $y$ covered by phase $B$. Each phase has its own electron density $n_{A,B}$ and $ (1-y) n_A + y n_B=n$. Minimization of $U^{PS}$ according to $n_{A,B}$ leads to conditions for the boundaries of the phase separated area\cite{bak}:
\begin{equation}
 \left. \frac{\partial U^A}{\partial n_A}\right\vert_{n_1}= \left. \frac{\partial U^B}{\partial n_B}\right\vert_{n_2}\label{maxw}
\end{equation}
Between $n_1$ and $n_2$ the phases $A$ and $B$ coexist (cf. Fig. \ref{ps_gen}).

\section{\label{results}Results}

In this section we will present the numerical results of the self-consistent calculations. The effect of the different extensions will be studied in detail. All results are given for the spin $S=\frac{3}{2}$.

\subsection{\label{CI} Coulomb Interaction}

\begin{figure*}[tb]

\begin{minipage}{.3\linewidth}
	 \includegraphics[width=\linewidth]{dos_mm_J0.75W1g0.eps}
\end{minipage}
\begin{minipage}{.03\linewidth}
\ 
\end{minipage}
\begin{minipage}{.3\linewidth}
	 \includegraphics[width=\linewidth]{dos_oU_J0.75W1g0.eps}
\end{minipage}
\begin{minipage}{.03\linewidth}
\ 
\end{minipage}
\begin{minipage}{.3\linewidth}
	 \includegraphics[width=\linewidth]{dos_mU_J0.75W1g0.eps}
\end{minipage}

\begin{minipage}{.3\linewidth}
	 \includegraphics[width=\linewidth]{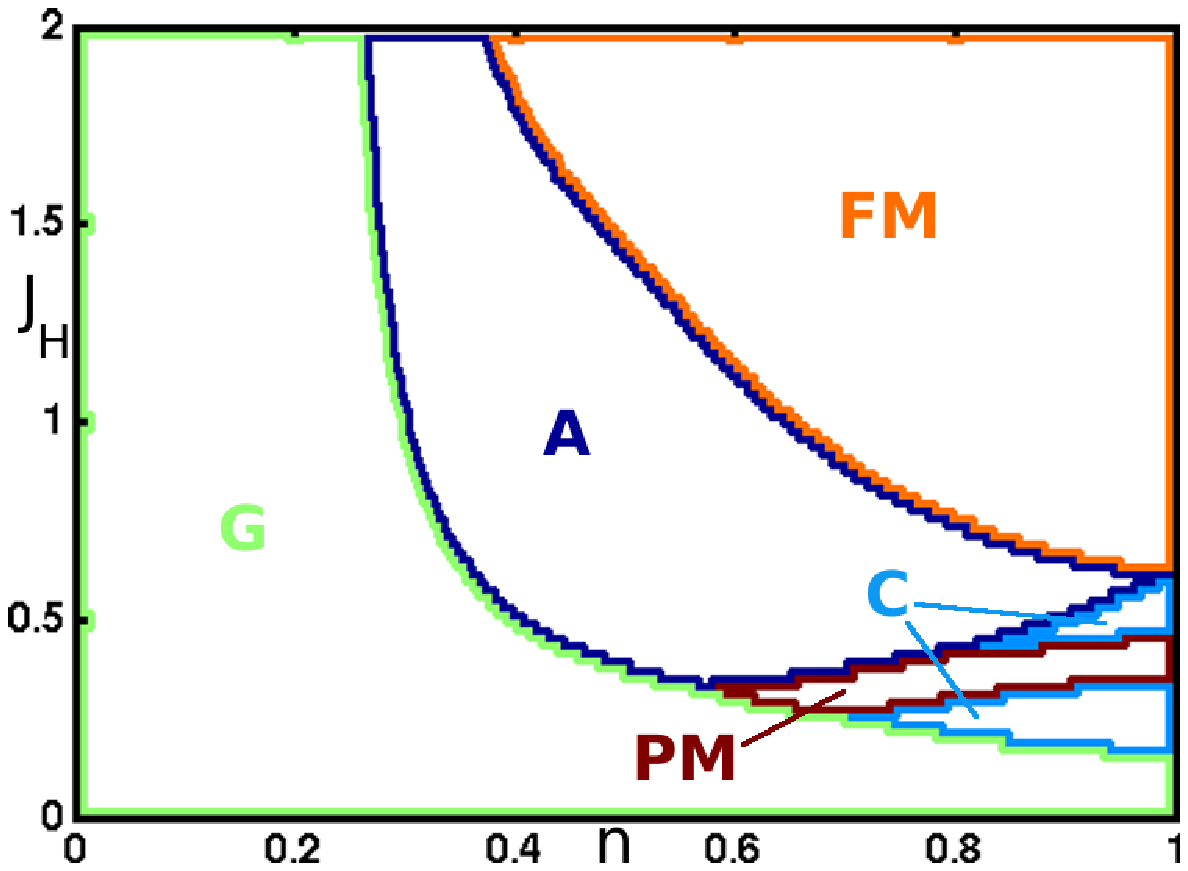}
\end{minipage}
\begin{minipage}{.03\linewidth}
\ 
\end{minipage}
\begin{minipage}{.3\linewidth}
	 \includegraphics[width=\linewidth]{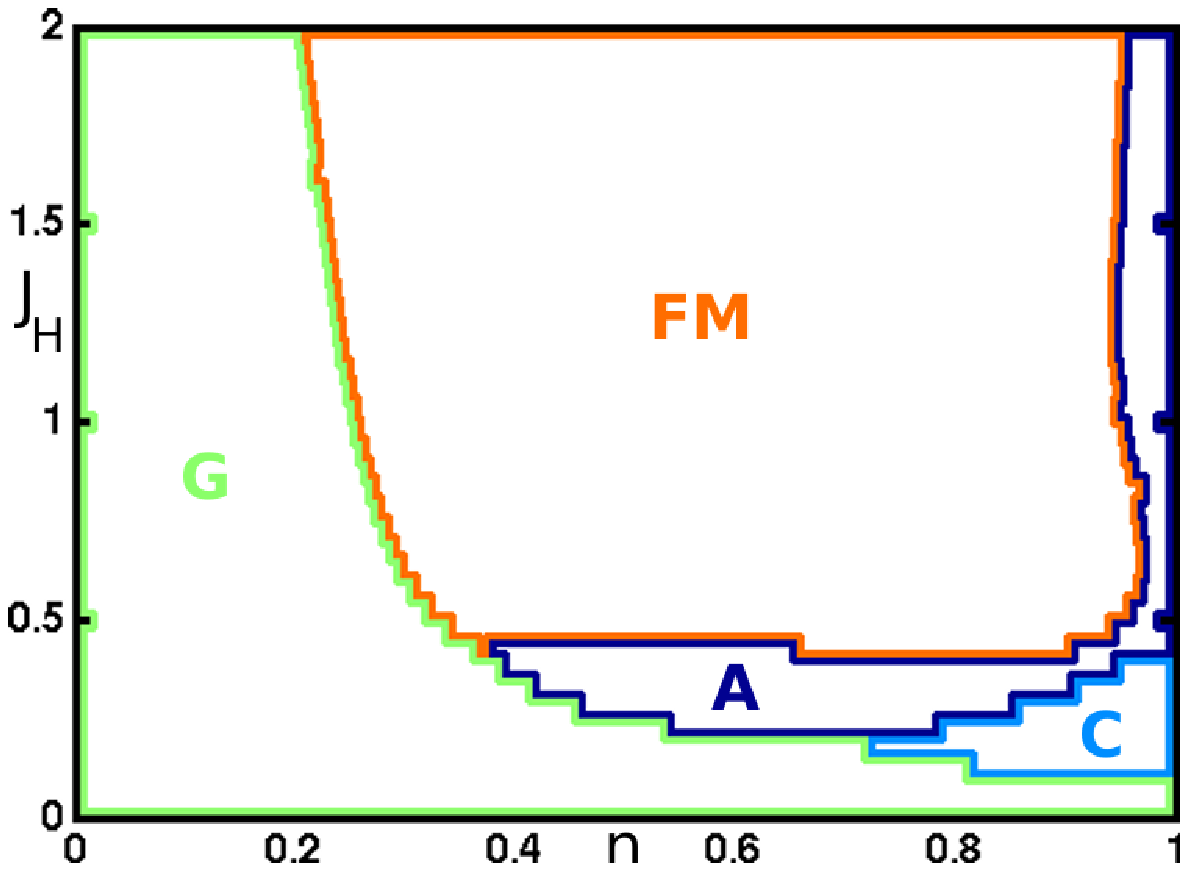}
\end{minipage}
\begin{minipage}{.03\linewidth}
\ 
\end{minipage}
\begin{minipage}{.3\linewidth}
	 \includegraphics[width=\linewidth]{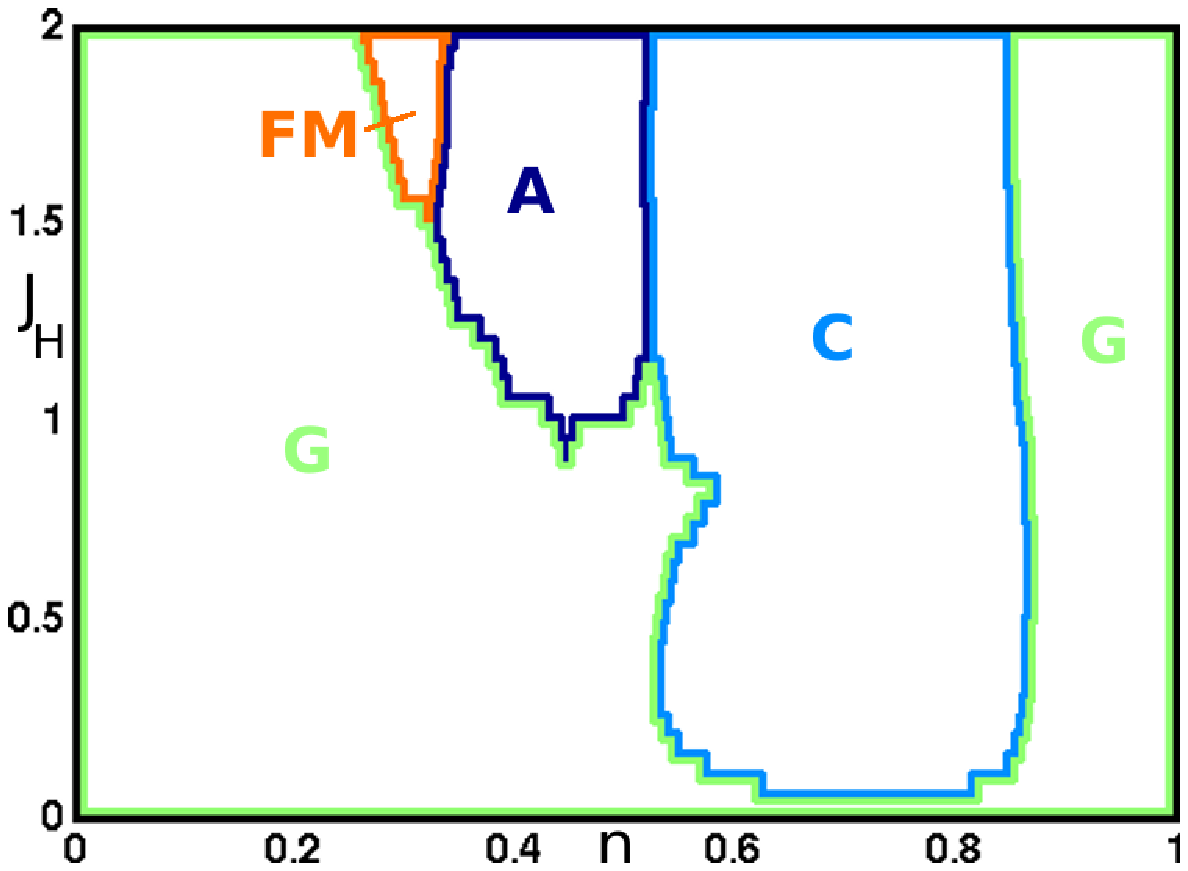}
\end{minipage}

\caption{\label{pd_diff_coul}(color online) \textit{upper line:} QDOS and chemical potential (vertical lines) at $n=0.94$ for ferromagnetic (solid black line) and A-type antiferromagnetic (dashed red line) order at $J_H=0.75$eV. \textit{left:} No Coulomb repulsion, \textit{middle:} only intraband repulsion, \textit{right:} intra+interband repulsion \textit{all:} $W=1$eV, $J_{AF}=1.7$meV, $g=0$. \textit{lower line:} Phase diagrams in dependence of the Hund's coupling $J_H$ in eV and the electron density $n$ for different types of magnetic order. For vanishing antiferromagnetic coupling $J_{AF}$ the whole phase diagram for intraband repulsion would be ferromagnetic and also everything below $n\approx 0.94$ for intra+interband repulsion.}
\end{figure*} 

Basically the Coulomb repulsion can act in three ways. It can be turned off completely, act only between electrons in the same band or between electrons of different bands, too. In our treatment it affects the bandwidth and the spectral weight of each band. The original values are reduced by the factor $\gamma_{\alpha\sigma}$ in (\ref{specw}). Because these factors are dependent on the occupation number the main differences occur at high electron densities. For example at $n=1$ the bands are quarter-filled if there is no Coulomb repulsion, one-third-filled at intraband repulsion and completely filled with additional interband repulsion. Ferromagnetism at large couplings $J_H$ in the pure double-exchange model is most favored at quarter-fillings \cite{mc_de}. This is the maximum filling that can be reached with an absent Coulomb interaction for densities $0\leq n\leq 1$. On the other hand there is no ferromagnetism in the DE model at half-filling where the chemical potential lies between the bands. A comparable situation in our extended KLM can only be achieved with intra+interband Coulomb repulsion at $n\approx 0.94$ for (anti)ferromagnetic phases (Fig. \ref{pd_diff_coul}). So one can explain basic differences of the phase diagrams in Fig. \ref{pd_diff_coul}.\\
Without Coulomb interaction we have effective quarter-filling at $n=1$. This means that ferromagnetism is mostly provided by the itinerant electrons and the energy difference $\Delta U_{el} = U_{el}^{AFM}- U_{el}^{FM}$ has its maximum at this electron density. With a finite antiferromagnetic coupling $J_{AF}$ antiferromagnetic phases begin to appear at lower electron densities $n$ where the $\Delta U_{el}$ are small (cf. Fig.  \ref{deltaU_dCI},\ref{deltaU}). Thus the total energy differences are mainly governed by the energy of the localized spins (\ref{uss}) which is density-independent. The G-type AFM gains most energy from the direct antiferromagnetic coupling and exists at the lowest electron densities (compare Section \ref{daf}).\\
If we have intraband repulsion antiferromagnetic phases also appear at higher densities. Because of the reduction of the total spectral weight we cross the effective quarter-filling and the absolute energy difference of the phases is reduced again (cf. Fig. \ref{deltaU_dCI}). Therefore we get antiferromagnetic phases if we have a sufficient antiferromagnetic coupling $J_{AF}$ at $n=1$ without losing the ferromagnetic phase for $n<1$. This was not possible without Coulomb interaction where the ferromagnetic phase stays longest at $n=1$.\\
When we also add the interband repulsion the spectral weight will be reduced even more. Thus the maximum absolute electronic energy difference for the ferromagnetic phase lies at $n\approx 0.4$ (cf. Fig. \ref{deltaU}). There the ferromagnetic phase exists longest with increasing $J_{AF}$ (cf. Fig. \ref{pd_diff_coul}). As a second feature the upper subband will be occupied for electron densities $n\approx 0.94$ (resp. $n\approx 0.91$ for the paramagnetic phase). Because of the extremely reduced spectral weight the energy differences become very small and we get many small regions of different phases for $n\geq 0.94$ at $J_{AF}=0$. This occurrence of many small (anti-) ferromagnetic regions could be a hint for phase separation or spin-canted states. We will discuss phase separation in section \ref{ps_res}. With finite $J_{AF}$ the interval $0.94\lesssim n \leq 1$ is dominated by the energy of the localized spins and that is why the G-type AFM appears.\\
In general the Coulomb interaction favors ferromagnetic order. For both types of repulsion  the whole phase diagram would be ferromagnetic if $J_{AF}=0$ (except at $n\approx 0.94$ at interband repulsion) even for very low $J_H$ and $U$  (Ref. \cite{mm}). It can be argued that the interplay between the two ordering mechanisms, the Coulomb interaction and the double exchange, makes it easier for the system to go to the symmetric state. But the reduction of the spectral weight and the bandwidth lower the absolute energy differences of the itinerant electron system. Thus one needs a smaller $J_{AF}$ to create antiferromagnetic phases. For this reason the interband Coulomb phase diagram in Fig. \ref{pd_diff_coul} has larger regions of antiferromagnetic phases compared to the others.\\
From now on we will only discuss the cases of intra or intra+interband Coulomb repulsion.

\subsection{\label{hund} Hund's Coupling}

\begin{figure*}[tb]

\begin{minipage}{.3\linewidth}
	 \includegraphics[width=\linewidth]{dos_mU_J0.05W1g0.eps}
\end{minipage}
\begin{minipage}{.03\linewidth}
\ 
\end{minipage}
\begin{minipage}{.3\linewidth}
	 \includegraphics[width=\linewidth]{dos_mU_J0.15W1g0.eps}
\end{minipage}
\begin{minipage}{.03\linewidth}
\ 
\end{minipage}
\begin{minipage}{.3\linewidth}
	 \includegraphics[width=\linewidth]{dos_mU_J0.3W1g0.eps}
\end{minipage}

\begin{minipage}{.3\linewidth}
	 \includegraphics[width=\linewidth]{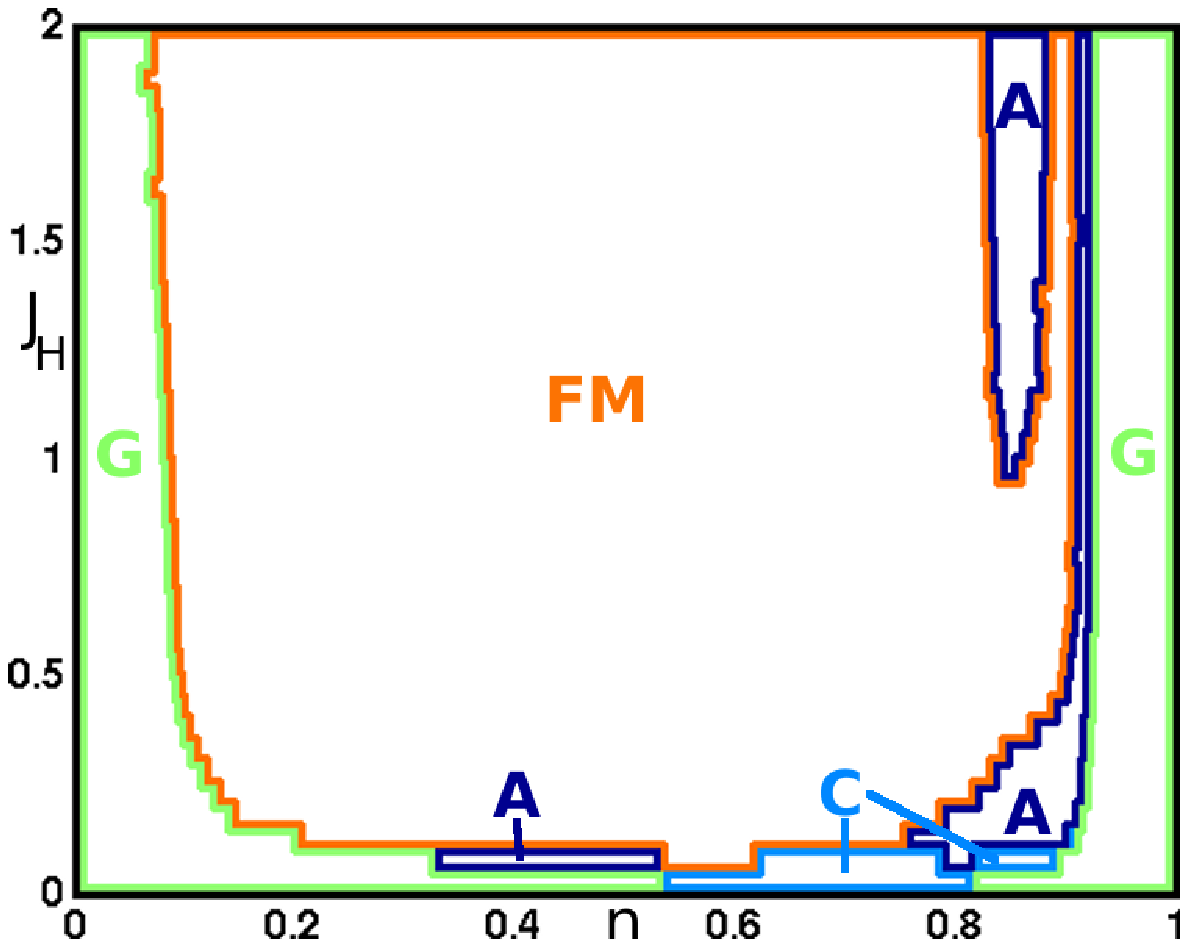}
\end{minipage}
\begin{minipage}{.03\linewidth}
\ 
\end{minipage}
\begin{minipage}{.3\linewidth}
	 \includegraphics[width=\linewidth]{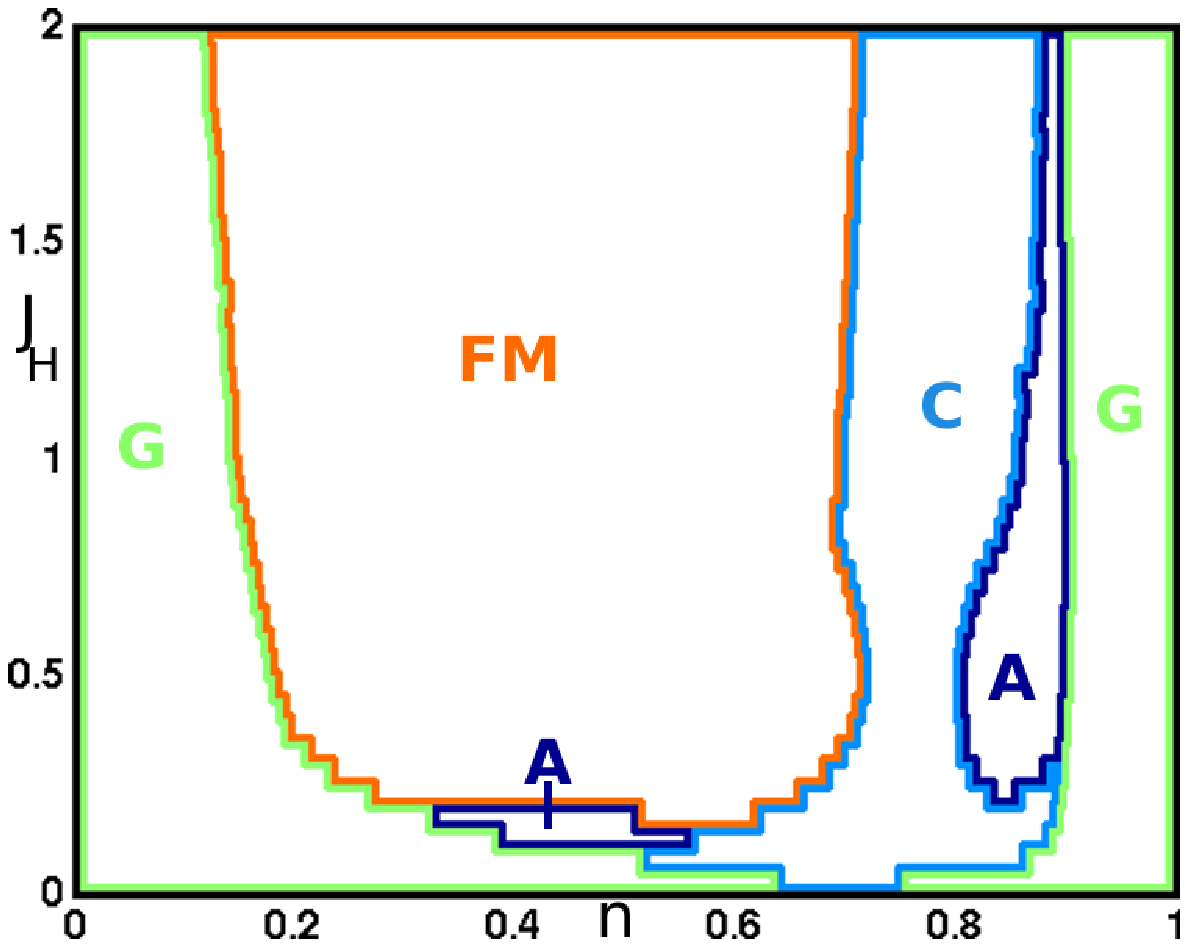}
\end{minipage}
\begin{minipage}{.03\linewidth}
\ 
\end{minipage}
\begin{minipage}{.3\linewidth}
	 \includegraphics[width=\linewidth]{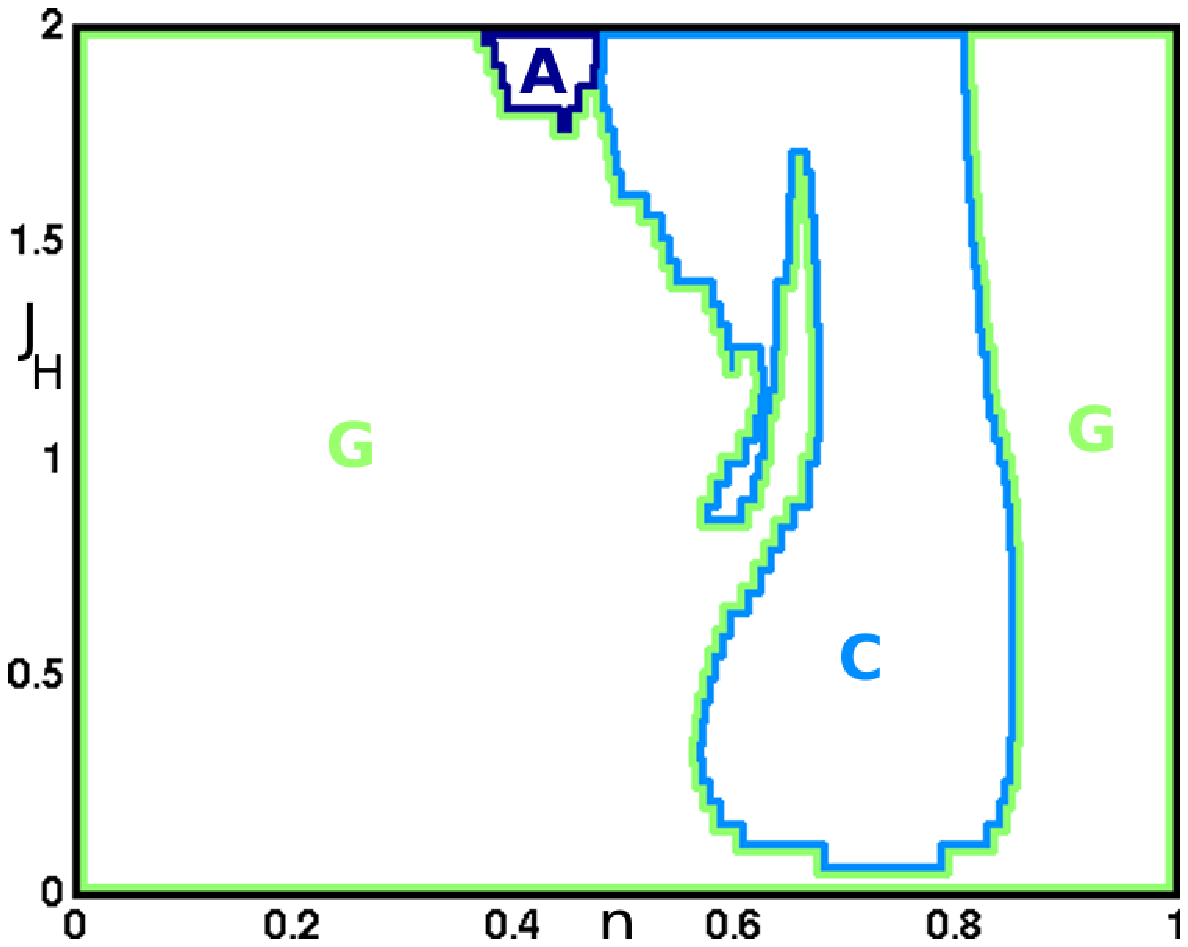}
\end{minipage}

\caption{\label{pd_jn_diff_jaf}(color online) \textit{upper line:} QDOS and chemical potential for different Hund's couplings $J_H$ and $n=0.9$. Solid (black) line ferromagnetism, dashed (red) line A-type AFM.  \textit{lower line:} Phase diagrams in dependence of the Hund's coupling $J_H$ in eV and the electron density $n$ for different types of magnetic order. \textit{left:} $J_{AF}=0.56$meV, \textit{middle:} $J_{AF}=0.94$meV, \textit{right:} $J_{AF}=1.81$meV \textit{all:} intra+interband Coulomb repulsion $W=1$eV, $g=0$.}
\end{figure*}

In the ferromagnetic KLM the Hund's coupling $J_H$ in principle favors ferromagnetism. But because of the complex interplay of a parallel alignment of the localized/itinerant spins and the hopping there also occur antiferromagnetic phases. These exist mainly at low and intermediate couplings $J_H$ and at high electron densities\cite{zt_pd}. This difference between low and strong coupling has its origin in the splitting of the spin-up and spin-down bands (Fig. \ref{pd_jn_diff_jaf}). The quasi-particle DOS changes strongest at low couplings and there are more spin-down states occupied. At some $J^c_H$ the main part of the spin-down band is shifted above the chemical potential depending on the electron density $n$. For strong couplings there is no major variation of the shape of the QDOS. Only the splitting gets larger with larger $J_H$. For example this is reflected in the Curie temperature which cannot be increased by increasing $J_H$ at large couplings\cite{ms2}.\\
Since we have an additional Hubbard part to our basic KLM we have two supporting mechanisms that create ferromagnetism. This means we would have an almost complete ferromagnetic phase diagram if we would choose $J_{AF}=0$. With a finite $J_{AF}$ one can see the emergence of different antiferromagnetic phases. Figure \ref{pd_jn_diff_jaf} shows the disparity between low and strong coupling. For low $J_H$ the boundaries between the phases are dependent on the Hund's coupling and the electron density. On the other hand, at strong couplings, these boundaries occur at constant densities $n_c$ and therefore they are almost vertical in the phase diagram. The phases are now very stable at certain densities $n$ concerning a change of $J_H$. This stability vanishes with increasing $J_{AF}$  when the electronic energy differences become to small.\\
The main effect of $J_{AF}$ happens of course at small differences of the electron energies (Fig. \ref{deltaU}). That is given, for intra+interband Coulomb interaction, at small and high $n$. A small $n$ means that there are not enough electrons to create large absolute energy differences. On the other hand a large $n$ leads to an occupation of the upper sub-band (cf. \ref{CI}) which is at higher energies. For $n\rightarrow 1$ the electronic internal energies for all phases go to zero and therefore the energy differences vanish, too.

\subsection{\label{daf} Direct Antiferromagnetic Coupling}

\begin{figure*}[tb]
\begin{minipage}{.31\linewidth}
	 \includegraphics[width=\linewidth, height=4.5cm]{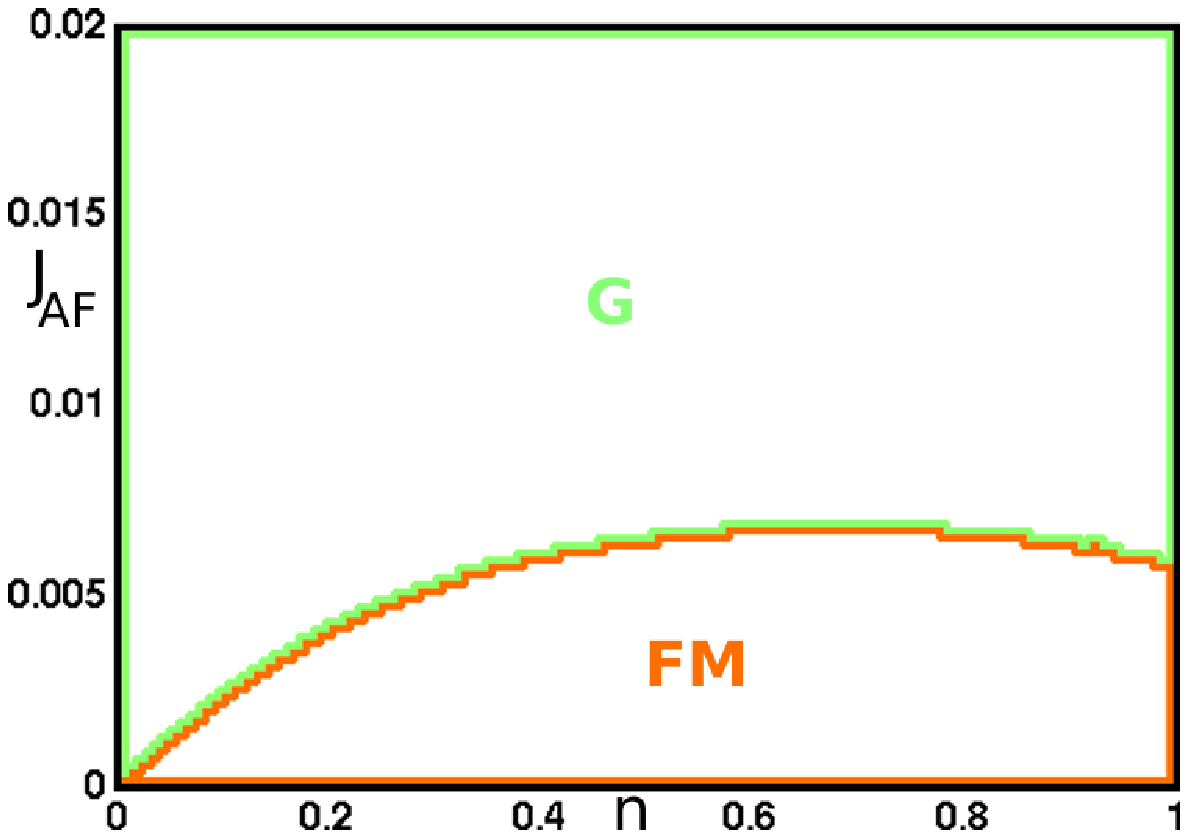}
\end{minipage}
\begin{minipage}{.01\linewidth}
\ 
\end{minipage}
\begin{minipage}{.31\linewidth}
	 \includegraphics[width=\linewidth, height=4.5cm]{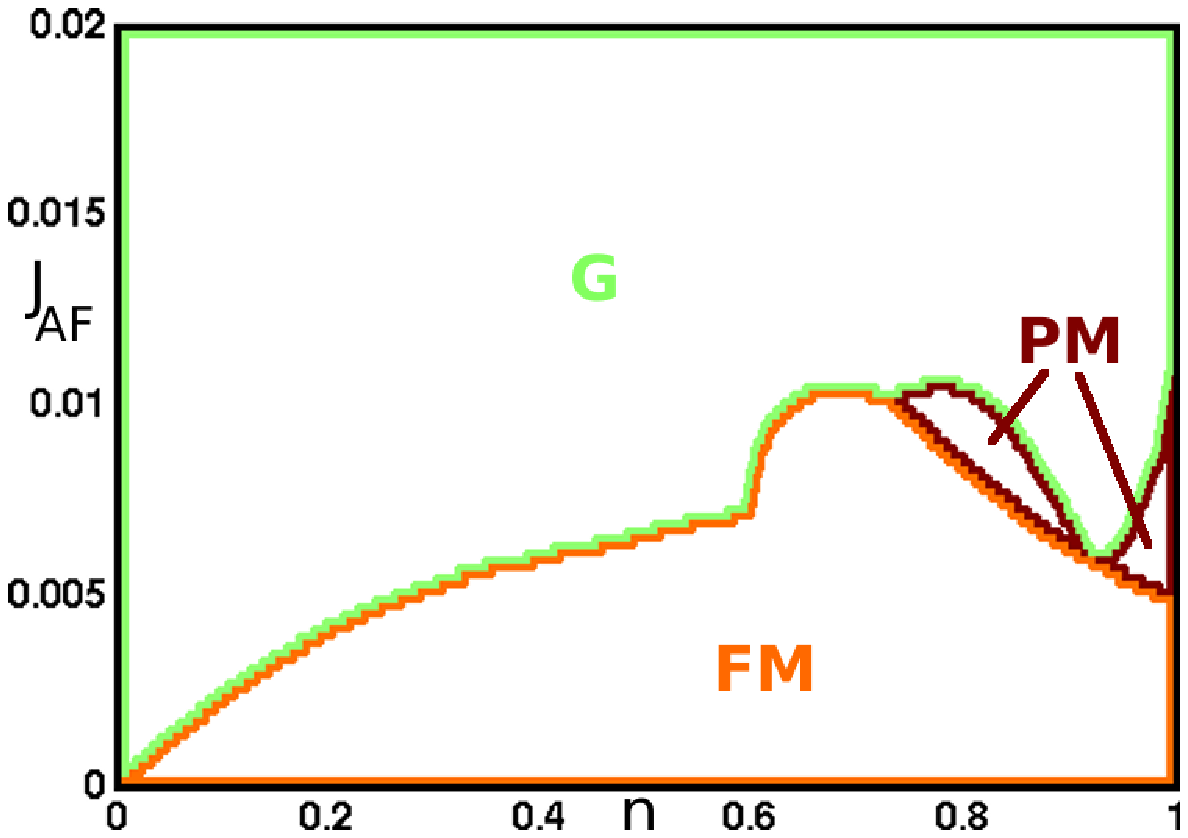}
\end{minipage}
\begin{minipage}{.01\linewidth}
\ 
\end{minipage}
\begin{minipage}{.31\linewidth}
	 \includegraphics[width=\linewidth, height=4.5cm]{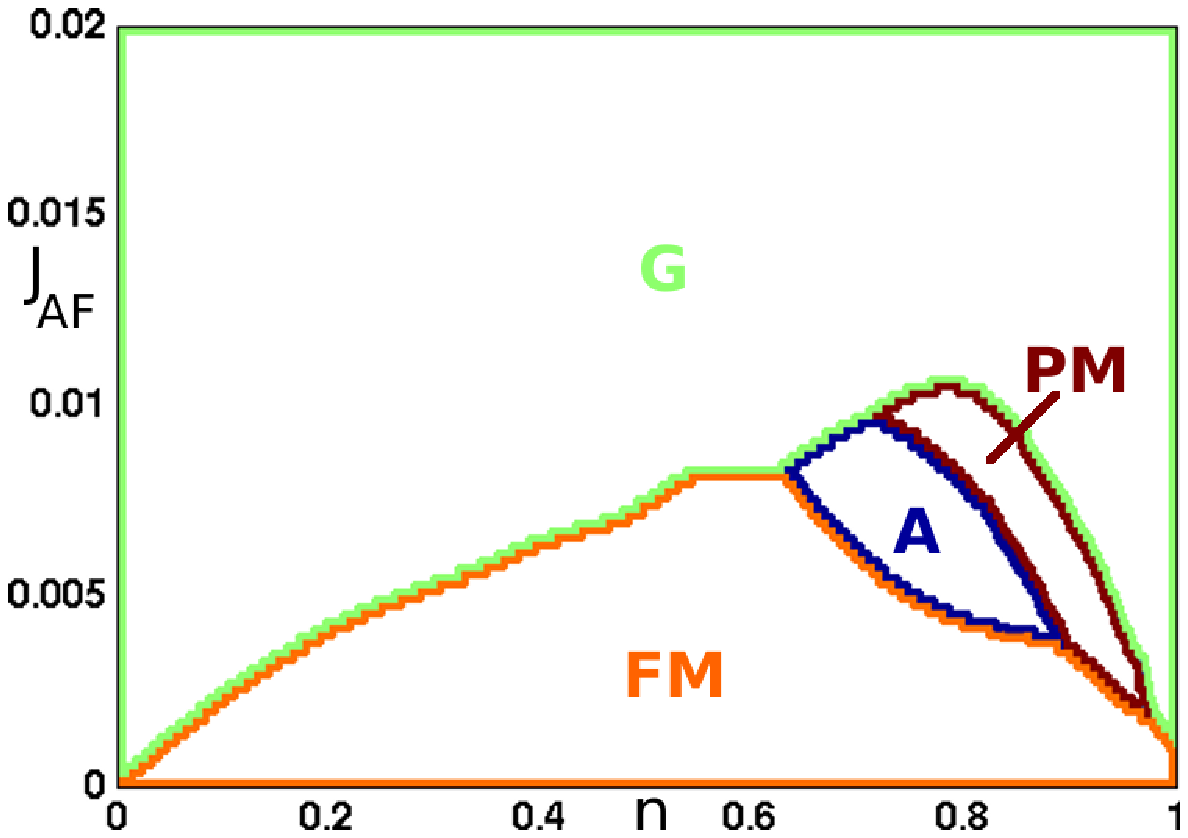}
\end{minipage}

\begin{minipage}{.31\linewidth}
	 \includegraphics[width=\linewidth, height=4.5cm]{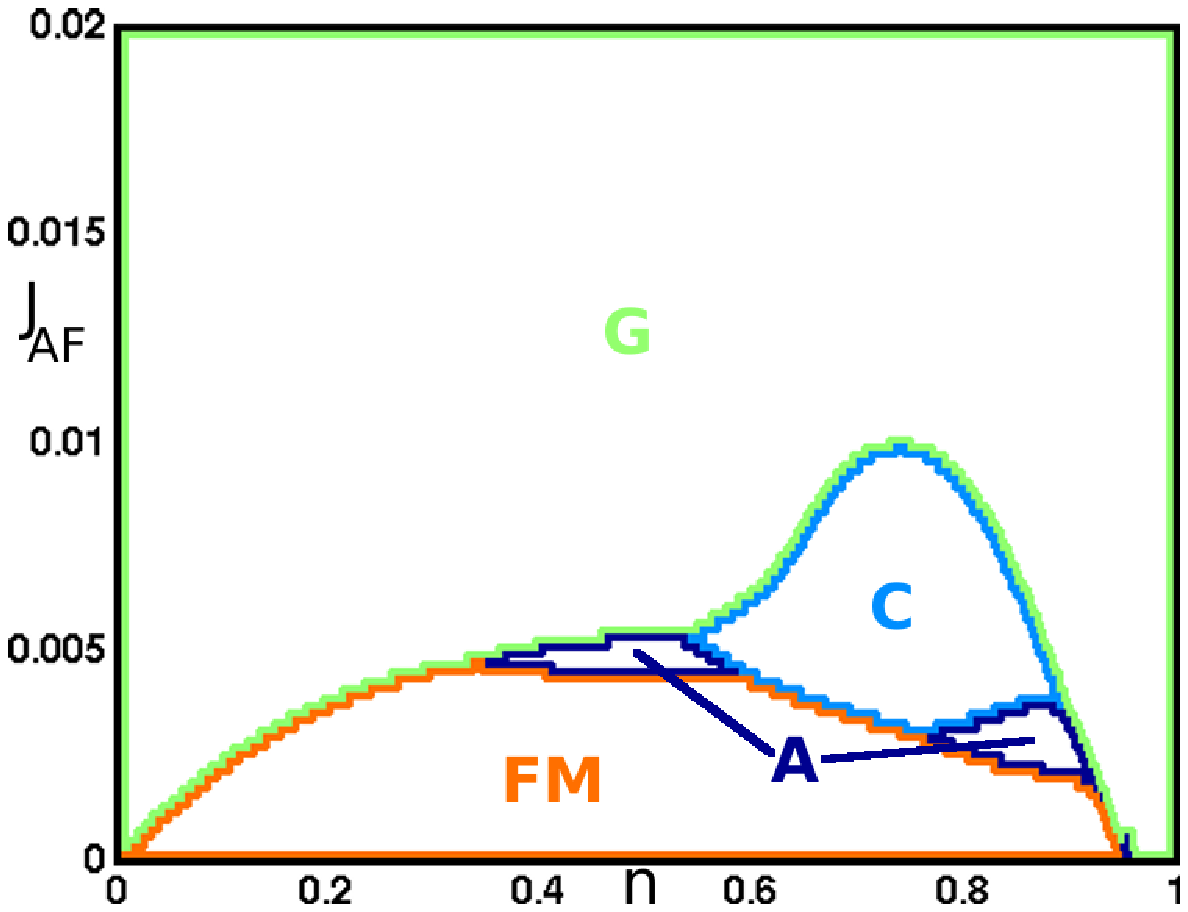}
\end{minipage}
\begin{minipage}{.01\linewidth}
\ 
\end{minipage}
\begin{minipage}{.31\linewidth}
	 \includegraphics[width=\linewidth, height=4.5cm]{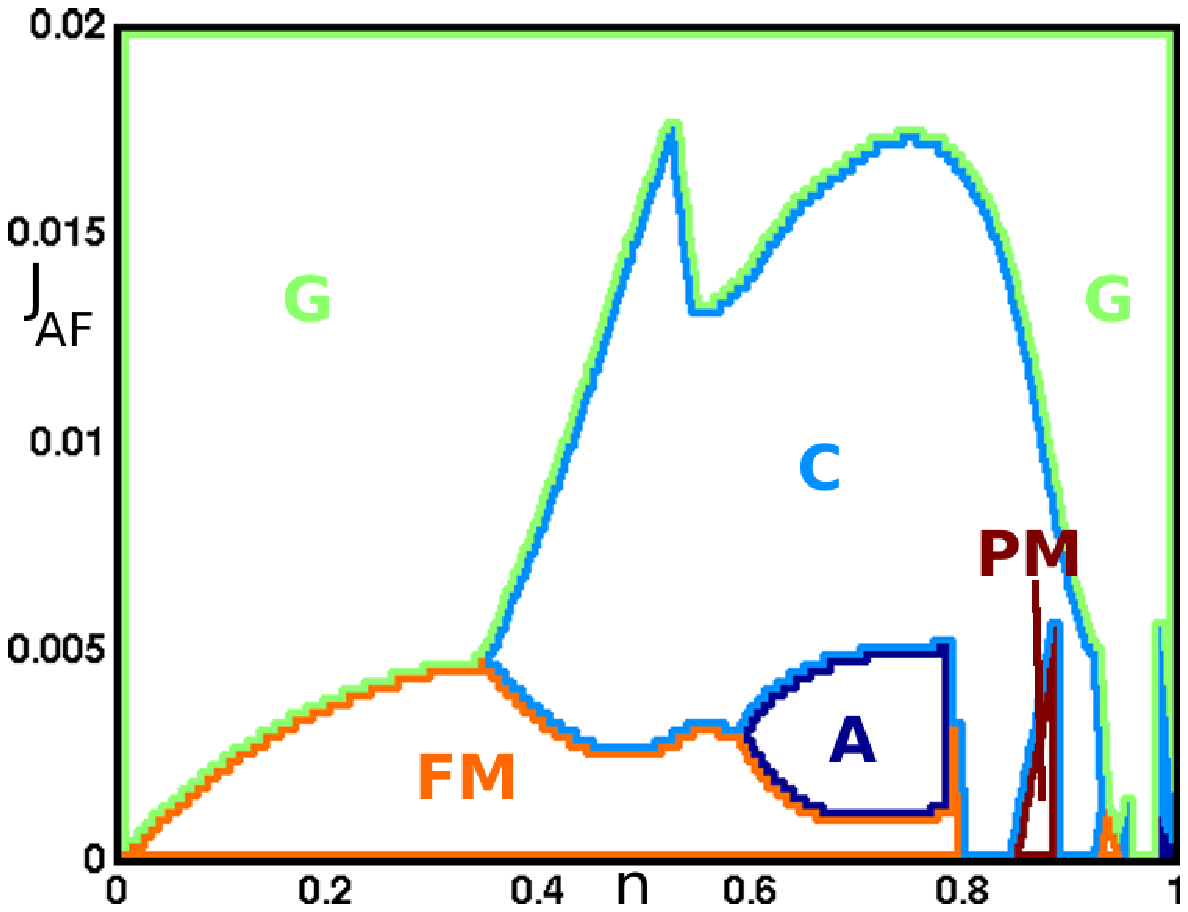}
\end{minipage}
\begin{minipage}{.01\linewidth}
\ 
\end{minipage}
\begin{minipage}{.31\linewidth}
	 \includegraphics[width=\linewidth, height=4.5cm]{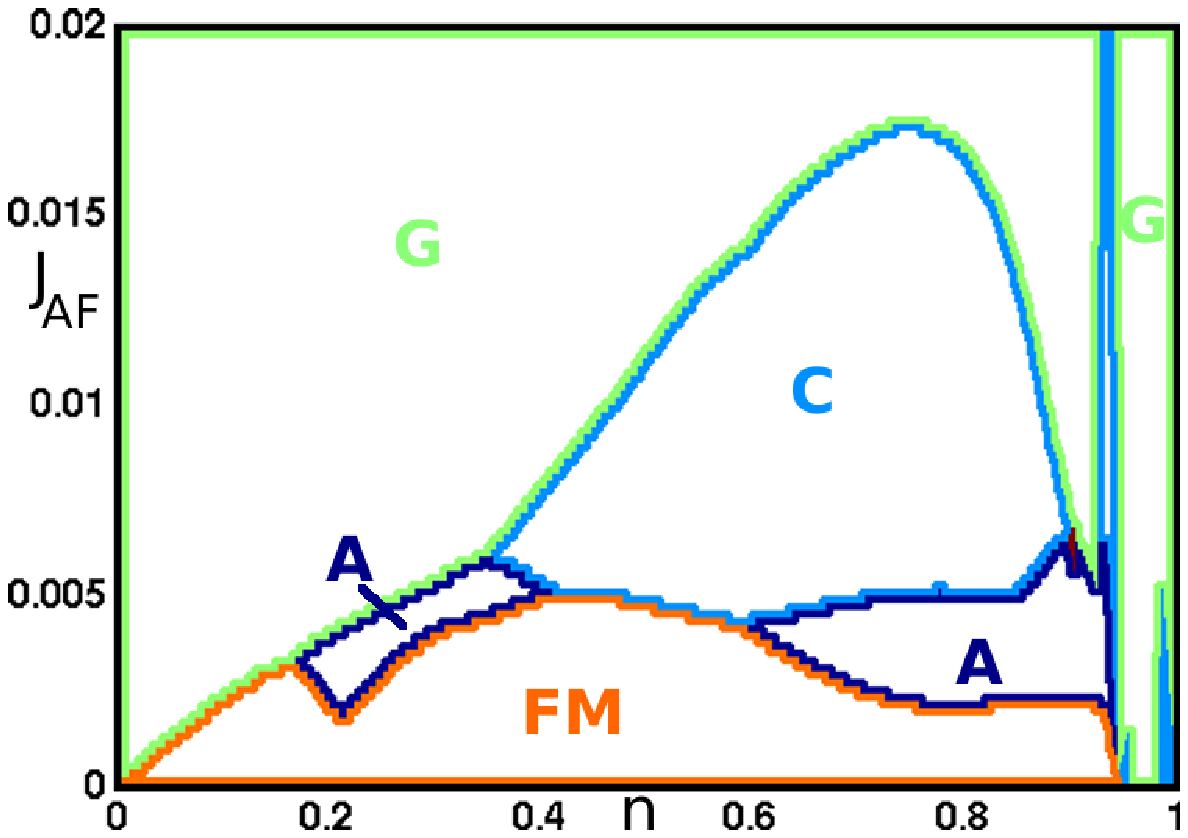}
\end{minipage}

\caption{\label{pd_jaf_n}(color online) Phase diagrams in dependence of the direct antiferromagnetic coupling $J_{AF}$ in eV and the electron density $n$ for different types of magnetic order. Additionally there is either no, intermediate or strong electron-phonon coupling $g$. \textit{upper line:} intraband Coulomb repulsion \textit{left:} $g=0\sqrt{eV}$, \textit{middle:} $g=0.9\sqrt{eV}$, \textit{right:} $g=1.0\sqrt{eV}$  \textit{lower line:} intra+interband Coulomb repulsion \textit{left:} $g=0\sqrt{eV}$, \textit{middle:} $g=0.6\sqrt{eV}$, \textit{right:} $g=1.0\sqrt{eV}$ \textit{all:}  $W=3$eV, $J_H=2$eV.}
\end{figure*}
 
In the last sections we investigated the influence of electronic correlations on the phase diagram. Their interplay mainly provided ferromagnetism. Thus we have already seen that we need a finite $J_{AF}$ to get larger regions of antiferromagnetic phases. Compared to the former mechanisms the direct antiferromagnetic coupling has no influence on the quasi-particle DOS. At finite temperatures it would of course act on the magnetization $\mean{S_z}$. But as we look at zero-temperature behavior we have restricted $\mean{S_z}$ to its maximum value. That is why $J_{AF}$ only influences the energy of the localized spins (\ref{uss}) and can be examined more separately.\\
The energy (\ref{uss}) is primarily dependent on the number of (anti)parallel ordered spins. As the G-type AFM has the most antiparallel neighbors it gains most from a finite $J_{AF}$.  Vice versa the ferromagnetic order is most suppressed by it. The other phases get an intermediate energy change depending on their structure (compare Fig. \ref{mag_types}). This energy change competes with the energy difference of the electronic subsystem. Thus the G-type emerges at low absolute energy differences $\Delta U_{el}$ (at $n\approx 0$ or $n\approx 1$ for intra+interband Coulomb repulsion) even for low $J_{AF}$. The A- and C-type typically begin at intermediate parameters $J_{AF}$ and $n$ where the absolute  $\Delta U_{el}$ is the largest (Fig. \ref{pd_jaf_n},\ref{deltaU}). The G-type normally is unpreferred by the itinerant electron system in this region.\\
It is not always possible to get all phases by varying $J_{AF}$. At some parameters sets only ferromagnetism and the G-type appear, for example. Thus it seems that complex phases diagrams, like those of the manganites, need the interplay of different interactions. More phases can appear if one has a finite electron-phonon coupling $g$ (Fig. \ref{pd_jaf_n}). As described in the next section this coupling has an unequal effect on the particular phases. The G-type can be reached everytime with a sufficient $J_{AF}$, of course.

\subsection{\label{jtc} Jahn-Teller coupling}

\begin{figure*}[tb]
\begin{minipage}{.47\linewidth}
	 \includegraphics[width=\linewidth, height=5.5cm]{jt_split2a.eps}
\end{minipage}
\begin{minipage}{.03\linewidth}
\ 
\end{minipage}
\begin{minipage}{.47\linewidth}
	 \includegraphics[width=\linewidth, height=5.5cm]{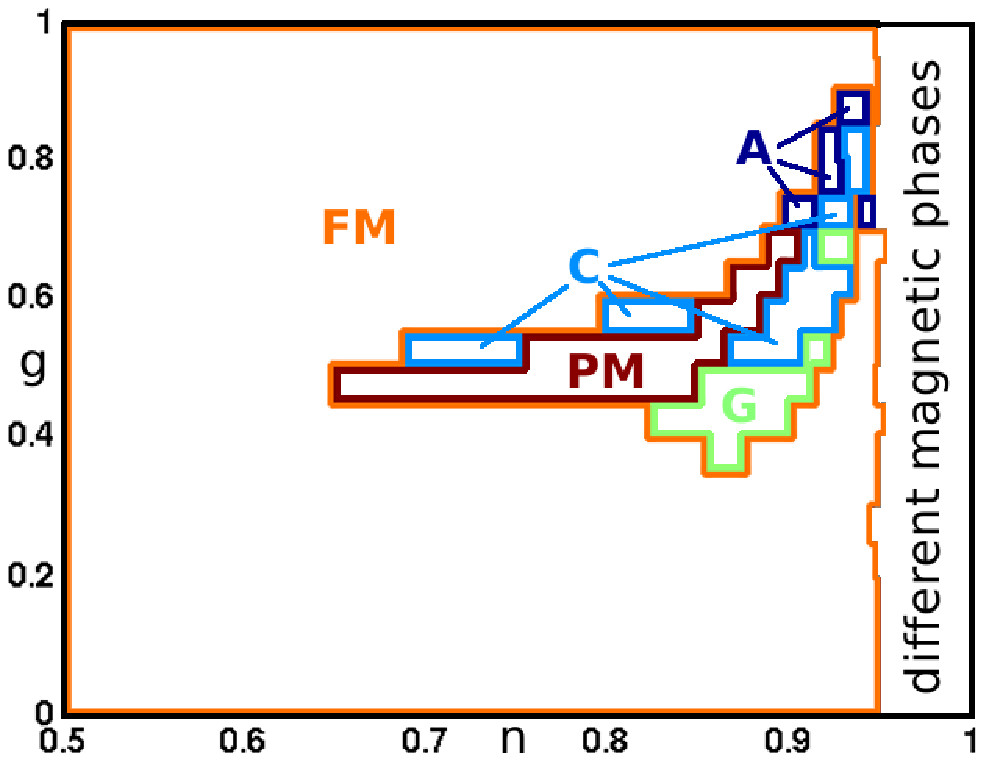}
\end{minipage}
\begin{minipage}{.47\linewidth}
	 \includegraphics[width=\linewidth, height=5.5cm]{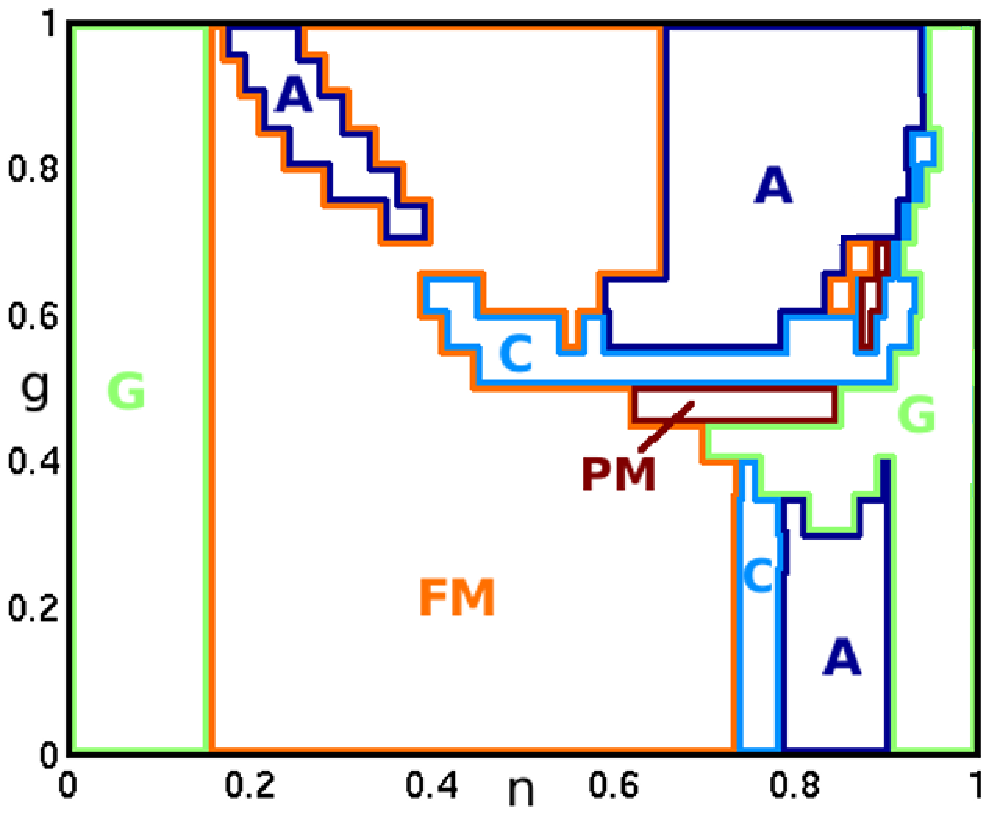}
\end{minipage}
\begin{minipage}{.03\linewidth}
\ 
\end{minipage}
\begin{minipage}{.47\linewidth}
	 \includegraphics[width=\linewidth, height=5.5cm]{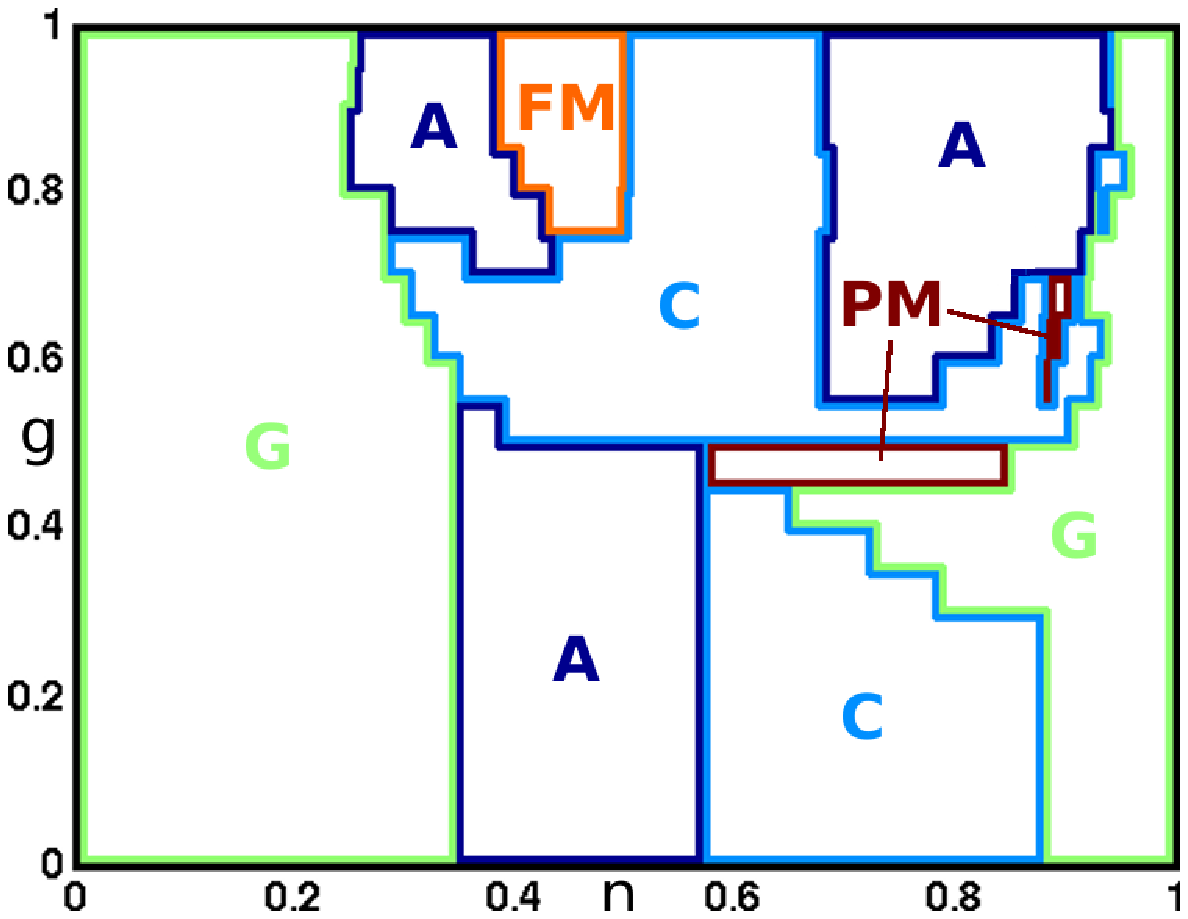}
\end{minipage}

\caption{\label{jt_split}(color online) Phase diagrams in dependence of the electron-phonon coupling $g$ in $\sqrt{\text{eV}}$ and the electron density $n$ for different types of magnetic order.  \textit{upper left:} Occurrence of the Jahn-Teller splitting.  The lines mark the beginning ($\mean{\Delta n} \approx \frac{1}{2} n$) of the splitting. \textit{remaining three graphs:} Favored magnetic phases for \textit{upper right:} $J_{AF}=0$ (mind the different $n$-scale!),\textit{lower left:} $J_{AF}=3$meV and \textit{lower right:} $J_{AF}=4.7$meV \textit{all:} $J_H=2$eV, $W=3$eV. The JTE lowers the energy. Thus new phases develop at the beginning of the JT splitting with increasing $J_{AF}$ and lead to a complex picture for intermediate couplings $g$. Values for $n\gtrsim0.94$ are left out in the upper line due to graphical reasons. }
\end{figure*} 

The Jahn-Teller theorem says that a symmetric crystal with degenerated states lowers its symmetry so that it reduces its energy. Thus all the magnetic phases should lower their energy when the two orbitals (\ref{orb}) split. But as long as we calculate the splitting (\ref{split}) self-consistently it is not clear which coupling $g_c$ is necessary to lift the degeneracy. Figure \ref{jt_split} shows indeed that the critical coupling is not the same for the single ordering types. For example at low couplings $g\gtrsim 0.3$ the G-type AFM appears first in the density interval $0.6\lesssim n \lesssim 0.9$ when we use intra+interband Coulomb interaction. On the other hand the A-type AFM shows up first at large $g$ and low densities. It is not only important which phase is JT-split but also how much a single phase profits from the splitting. The effect of the JT splitting on the electronic internal energies can be seen in Fig. \ref{deltaU}.\\
It was mentioned before that the phase diagrams with finite Coulomb interaction and vanishing $J_{AF}$ are almost purely ferromagnetic. But with an increasing coupling $g$ the G-type starts at $n\approx 0.8$ and the crystal gets G-type ordered. When one increases $g$ further on, the JT splitting starts in the paramagnetic phase and it is now preferred, even though the G-type is still split, too. With further increase of $g$ other phases also appear. The energy lowering due to the JT-part of the Hamiltonian (\ref{jtham}) depends on the occupation difference (\ref{split}). The maximum value of (\ref{split}) is $\mean{\Delta n}_{\text{max}}=n$. Thus the JT energy is only sufficient to create AFM phase at higher densities ($J_{AF}=0$).\\
If we use a finite $J_{AF}$ one sees that the single phases develop at the boundaries of their splitting zones with increasing $J_{AF}$. That is why the A-type AFM starts at low densities and large $g$ where it is the only split phase. In the strong coupling region ($2 g^2 n > W$)  the boundaries are not electron density dependent any more. Thus these boundaries get vertical similar to the large $J_H$ case in Sec. \ref{hund}. But the edges are shifted compared to the non-split region at low $g$.\\
This different behavior for either no, intermediate or strong couplings can also be seen in Fig. \ref{pd_jaf_n}. At low and strong $g$ we see no sharp edges in the phase diagrams which corresponds to slight changes in the electronic energy (\ref{uel}). This is because there is either no splitting or at large couplings the bands in all phases are split to their maximum value $2g^2n$. For intermediate couplings there are again different critical densities $n_c$ for each phase. That is why we get sharp edges at these values.

\subsection{Phase Separation \label{ps_res}}

\begin{figure*}[tb]

\begin{minipage}{.31\linewidth}
	 \includegraphics[width=\linewidth, height=4.5cm]{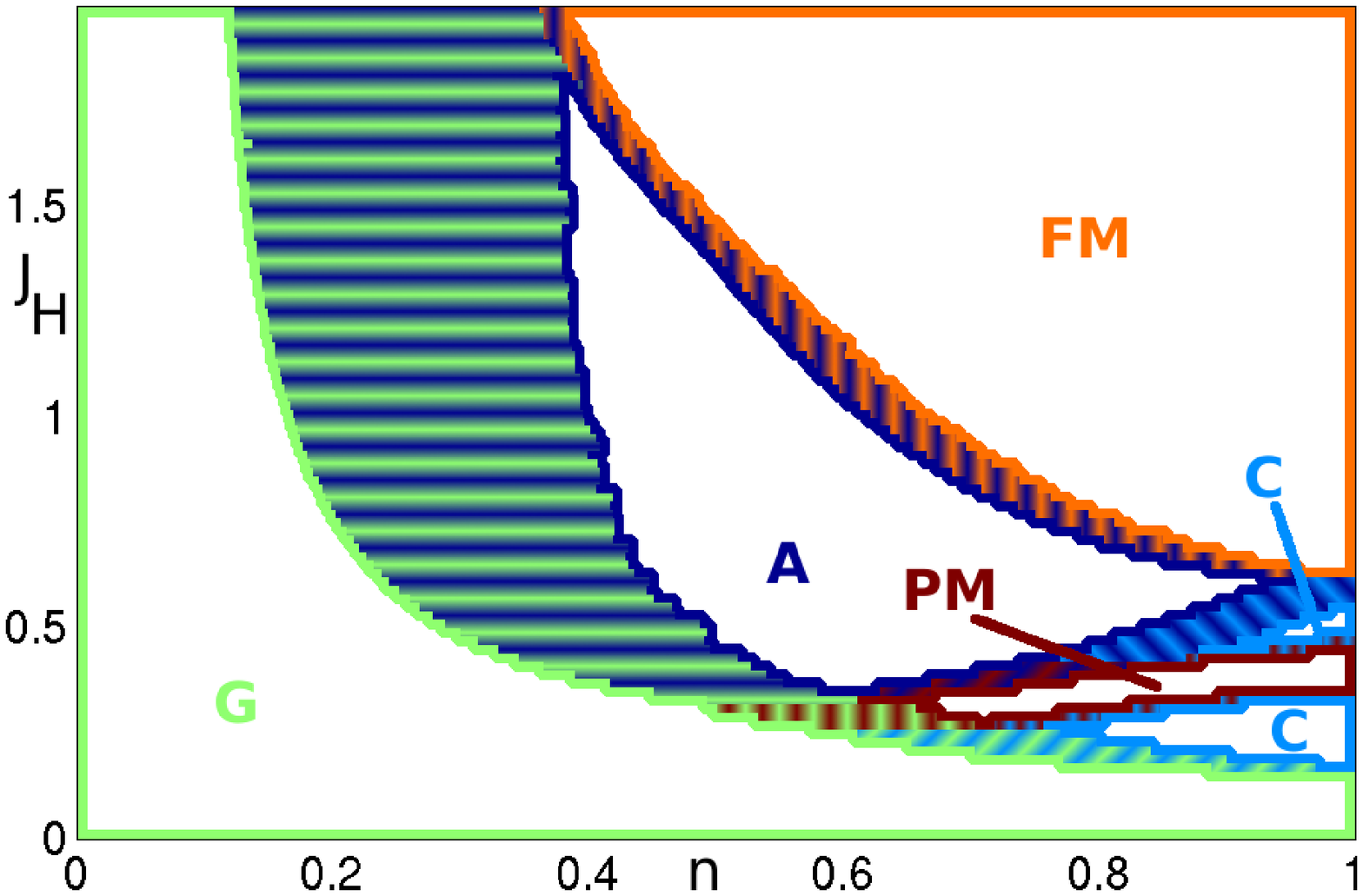}
\end{minipage}
\begin{minipage}{.01\linewidth}
\ 
\end{minipage}
\begin{minipage}{.31\linewidth}
	 \includegraphics[width=\linewidth, height=4.5cm]{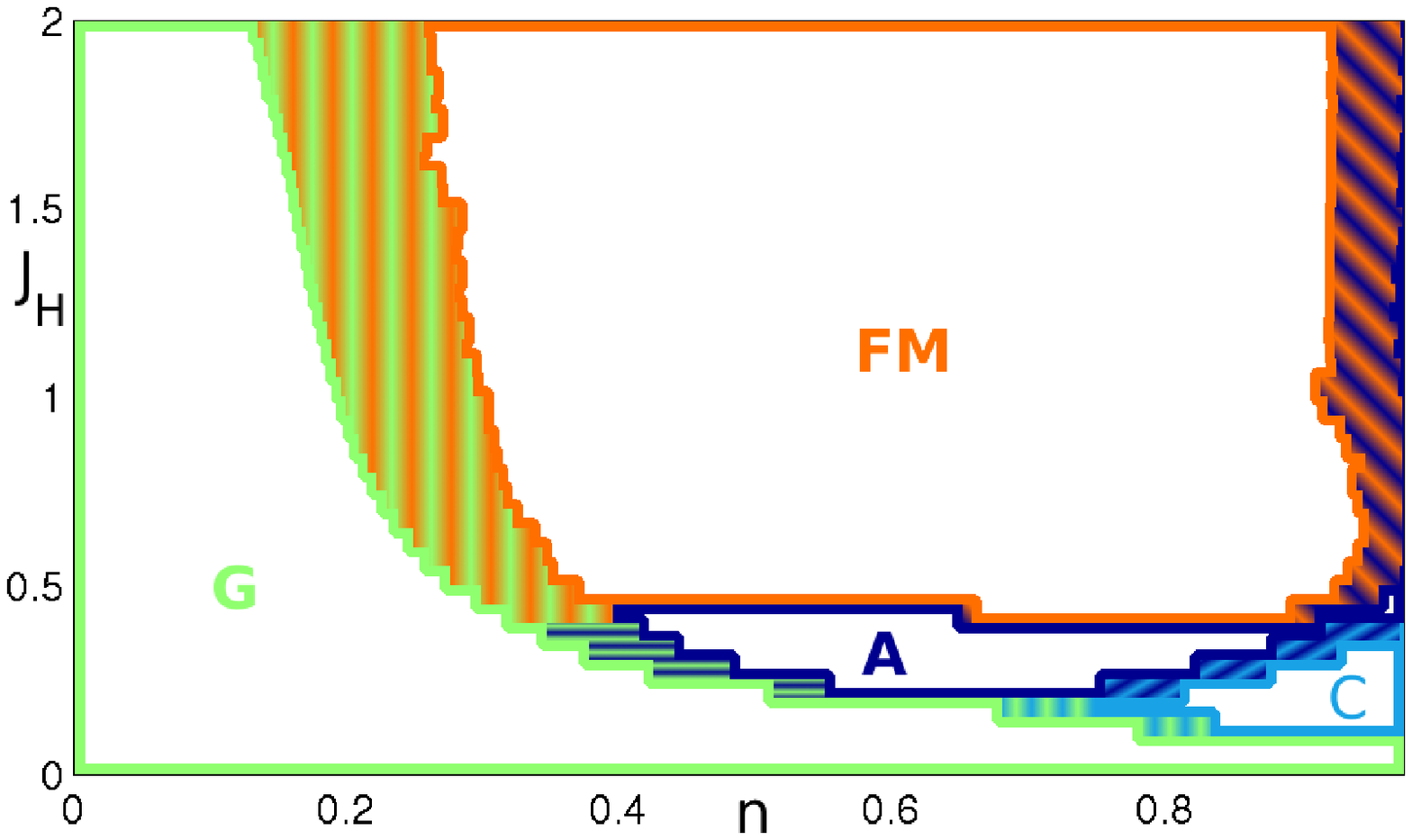}
\end{minipage}
\begin{minipage}{.01\linewidth}
\ 
\end{minipage}
\begin{minipage}{.31\linewidth}
	 \includegraphics[width=\linewidth, height=4.5cm]{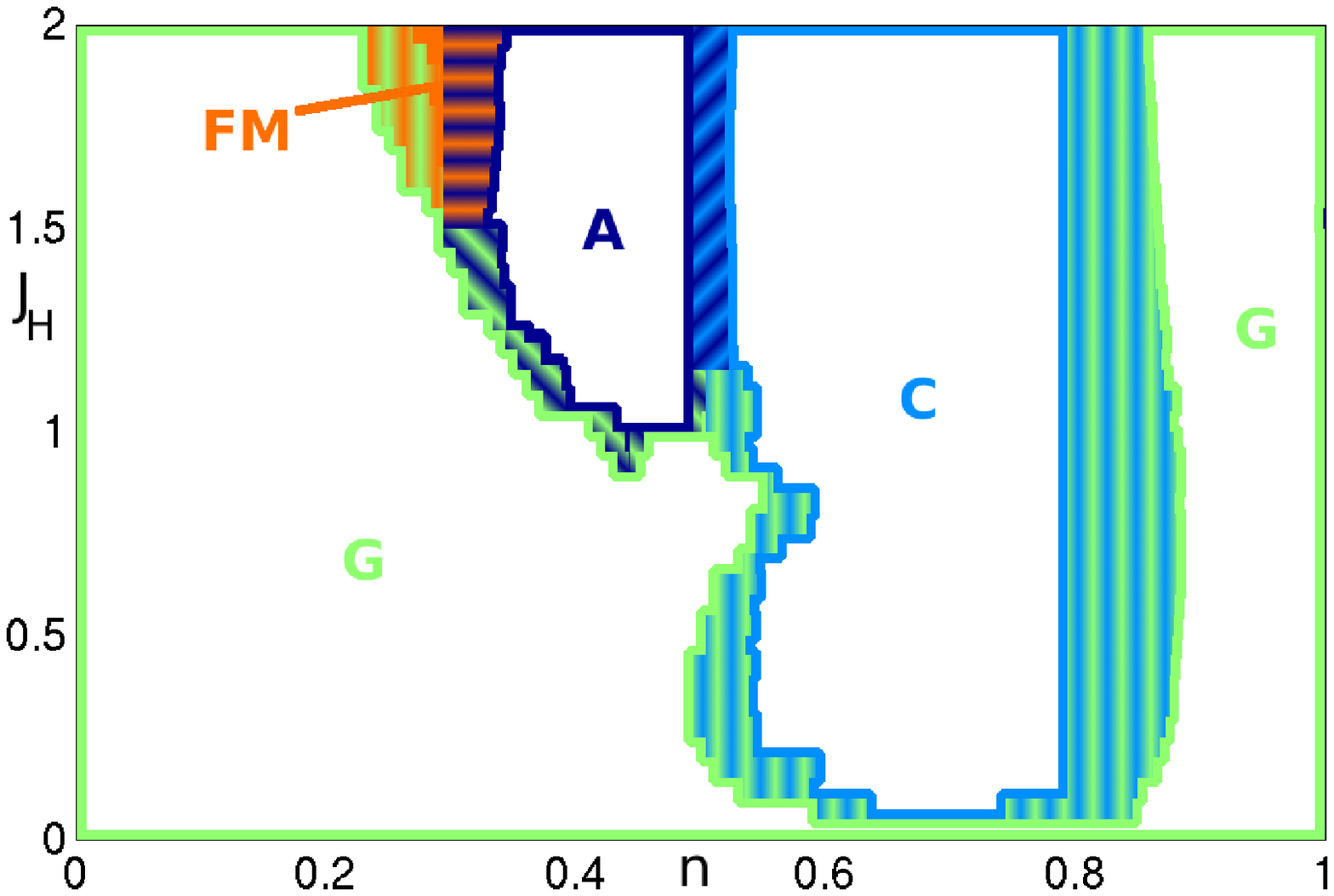}
\end{minipage}
\caption{\label{phase_sep}(color online) Phase diagrams of Fig. \ref{pd_diff_coul} but with the inclusion of phase separation (striped). The phase separation consists of the two phases which are adjacent to the left and right side of the phase separated region. \emph{left:} no Coulomb interaction \emph{middle:} intraband Coulomb repulsion \emph{right:} intra+interband Coulomb. Zones of phase separation become smaller with the increase of the influence of the Coulomb interaction. }
\end{figure*} 

In contrast to the homogeneous phases presented in the sections above, it is also possible that we have phase separated regions in the phase diagram. To see where a mixture of two phases can exist we use the method described in section \ref{ps}. Indeed we find that there is always a phase separated area between two phases. As an example we show in Fig. \ref{phase_sep} the regions of phase separation for the three different types of Coulomb interaction. If we have no Coulomb interaction we can get a very broad range of phase separation, especially at large $J_H$ and low $n$. With the increase of Coulomb repulsion partners the phase separation occurs only in smaller intervals around the original boundaries of the two phases. This has its reason in the gradient of the internal energy curves. If the gradient is small we get large regions of phases separation due to the construction we made in Eq. (\ref{maxw}). Figure \ref{deltaU_dCI} shows that we get the smallest gradient at low densities and the least influence of the Coulomb interaction.\\
It is also possible that complete homogeneous phases can vanish due to phases separation. For example, the ferromagnetic phase in Fig. \ref{phase_sep} at intra+interband Coulomb repulsion is, except for a small stripe, covered by phase separation of ferromagnetism/G-type antiferromagnetism or ferromagnetism/A-type respectively.\\
The inclusion of phase separation into the other phase diagrams leads to a qualitatively similar picture. Phase separation occurs always between two phases again in a more or less broad range. This would increase the variety of those phase diagrams even more and shows the complexity of the underlying model.

\section{\label{concl}Summary}

We have investigated the influence of several extensions of the two-band Kondo lattice model on the zero-temperature magnetic phase diagrams. Those extensions were a Hubbard term, direct antiferromagnetic coupling of the localized spins and electron-phonon coupling due to a Jahn-Teller part. We used self-consistent calculations, which go beyond the mean-field level. It has been seen that these extensions have a large impact on the phase diagrams. Because of the mutual effect on each other we can get very complex phase diagrams.\\
The Coulomb interaction normally leads to a ferromagnetic state. But due to a lowering of the spectral weight and bandwidth of the quasi-particle density of states it also lowers the absolute energy differences of the electronic subsystem. That is why the occurrence of antiferromagnetic phases needs a smaller antiferromagnetic coupling $J_{AF}$ at intra+interband Coulomb interaction compared to the case that one has intraband repulsion, only. Especially because of the strong Coulomb interaction we used in our work we always needed a finite $J_{AF}$ to create larger regions of antiferromagnetic phases. Big values of $J_{AF}$ always provide G-type AFM but intermediate values can lead to other phases, too.\\
The electron-phonon coupling $g$ can have a very subtle influence on the magnetic ordering. We have to differ again between low, strong and especially intermediate couplings. At low $g$ there is no splitting of the two bands and for large $g$ the important band occupation difference (\ref{split}) is saturated at $\mean{\Delta n}= n$. The intermediate couplings show a large variety of different phases.\\
Further on we looked on the appearance of phase separation. We found that there is indeed always a phase separated region between two homogeneous phases. Depending on the parameters this region can be very broad or very small. For special cases a originally homogeneous phase can be completely covered by phase separation.\\
To get complex phase diagrams, like for the manganites, extension to the ferromagnetic Kondo lattice model seem to be necessary, especially at large $J_H$ and Hubbard $U$. Here intermediate values of the direct antiferromagnetic and/or electron-phonon coupling play an important role to get a big diversity of magnetic phases. This could mean that approaches which are only valid in low or strong coupling regimes are maybe not appropriate to describe such material classes.\\
The magnetic order can surely be influenced by other effects. That could be orbital and charge order as well as the temperature, of course. These extensions would increase the complexity of the calculations much more and are left for later work.

\begin{appendix}

\section{\label{app}Electronic Energies}
\begin{figure}[tb]
\includegraphics[width=\linewidth]{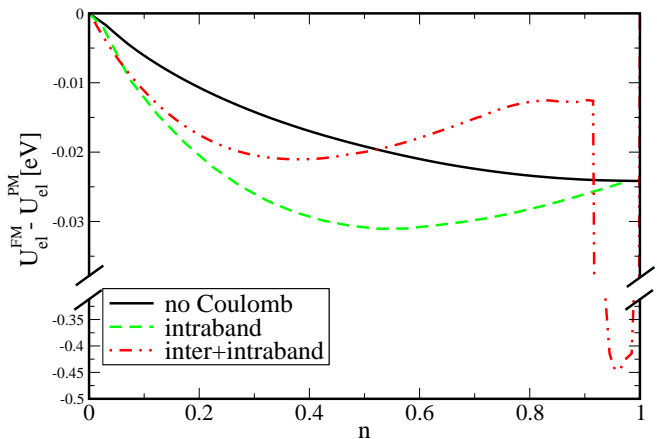}
\caption{\label{deltaU_dCI}(color online) Differences of the electronic internal energy (\ref{uel}) between ferromagnetic and paramagnetic phase for the single types of Coulomb interaction. The maximum of the absolute difference occurs at unequal electron densities $n$. At intra+interband repulsion the upper subband in the paramagnetic phase is occupied for $n\gtrsim 0.91$ and for the ferromagnetic phase at $n\gtrsim 0.94$. Thus there are larger energy differences in this region except for $n\rightarrow 1$ where the bands are completely filled and the energy difference goes to zero. Parameters: $J_H=1.5$eV, $W=1$eV, $g=0$}
\end{figure}
\begin{figure}[tb]
\includegraphics[width=\linewidth]{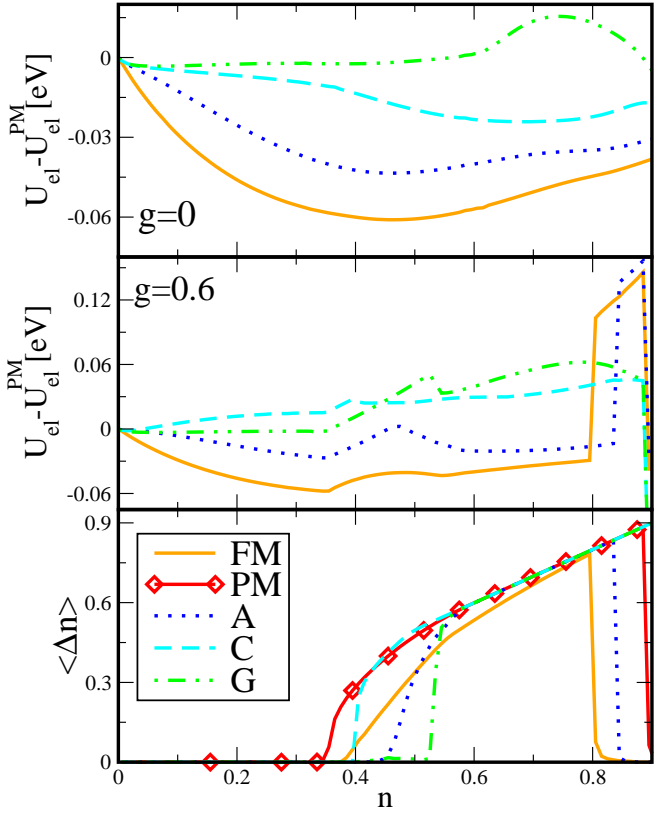}
\caption{\label{deltaU}(color online) \emph{Top and middle:} The electronic internal energy (\ref{uel}) differences between (anti)ferromagnetic phases and the paramagnetic phase at $g=0$ resp. $g=0.6\sqrt{eV}$ vs. the electron density $n$. \emph{Bottom:} The occupation difference $\mean{\Delta n}$ of the JT bands at intermediate coupling $g=0.6\sqrt{eV}$. - Low absolute energy differences at lower  $n$ lead  to a larger influence of the internal energy at finite $J_{AF}$ in this region. A finite (especially intermediate) JT coupling $g$ results in a drastic change of $\Delta U_{el}$. The JT splitting occurs at different electron density intervals for the single phases. For the given parameters the paramagnetic phase is split over the broadest range of $n$. Thus (anti)ferromagnetic phases are only possible if they also gain energy from the JT splitting, at least at higher $n$. Parameters: $J_H=2$eV, $J_{AF}=0$, $W=3$eV, intra+interband Coulomb repulsion}
\end{figure}

In this section the differences of electronic internal energies (\ref{uel}) between the (anti)ferromagnetic phases and the paramagnetic phases are shown. It can be seen in Fig. \ref{deltaU_dCI} and \ref{deltaU} that the lowest absolute energy differences occur at low electron densities $n$. In our treatment of the internal energy of the local moment system (\ref{uss}) the whole internal energy would be equally shifted for all $n$ at a finite $J_{AF}$. Thus it is most dominating at lower $n$ for all types of Coulomb interaction. At intra+inter band Coulomb repulsion at $n\gtrsim0.91$ the paramagnetic phase is always unpreferred but the energy differences between the (anti)ferromagnetic phases are very small. Therefore the local moment internal energy is most important here, too. The paramagnetic internal energy is not affected by a finite $J_{AF}$, the ferromagnetic phase and the A-type AFM are shifted to higher energies and the C and G-type AFM to lower ones. Therefore the G-type is preferred in those regions. In contrast to that we have a maximum of the absolute energy difference at a special $n_{\text{max}}$. The position of this maximum is mainly dependent on the type of the Coulomb interaction. It is at $n=1$ for the case of vanishing Hubbard repulsion and gets lower for finite Coulomb interactions. Near the $n_{\text{max}}$ the ferromagnetic phase remains longest with increasing $J_{AF}$.\\
If we have a finite JT coupling $g$ the JT bands are able to split. Since this splitting is calculated self-consistently it occurs only for special electron densities depending on $g$ and the type of the magnetic ordering (cf. Fig. \ref{jt_split}). As the JT splitting reduces the energy in most cases, magnetic phases with a finite band occupation difference $\mean{\Delta n}$ are preferred. Figure \ref{deltaU} shows the occurrence of the splitting for the single phases. Ferromagnetic order, for example, is only possible below $n\approx 0.8$ where the ferromagnetic system gets a finite $\mean{\Delta n}$ for the given parameters. At lower $n$ the JTE is less important because the maximum value of $\mean{\Delta n}^{\text{max}}=n$ is lower and the system is less profiting from a splitting.
\vspace{.5cm}

\end{appendix}

\end{document}